\RequirePackage{fix-cm}
\documentclass[smallextended]{svjour3}       
\smartqed  
\usepackage{graphicx}
\usepackage{amssymb}
\usepackage{amsfonts}
\usepackage{hyperref}
\usepackage[misc,geometry]{ifsym} 
\begin{document}

\title{Random spherical hyperbolic diffusion 
}


\author{Phil Broadbridge~\and~Alexander~D.~Kolesnik\\ \and Nikolai Leonenko \and Andriy Olenko}


\institute{P. Broadbridge \at
              Department of Mathematics and Statistics, La Trobe University, Melbourne, VIC, 3086, Australia\\
                          \email{P.Broadbridge@latrobe.edu.au}           
           \and
          A. D. Kolesnik \at
              Institute of Mathematics and Computer Science,  Kishinev, 2028, Moldova\\
              \email{kolesnik@math.md}          
              \and
               N. Leonenko \at
               Mathematics Institute, Cardiff University,  Cardiff, CF24 4AG, UK \\
               \email{leonenkon@cardiff.ac.uk} 
              \and
             \Letter\ A. Olenko \at
               Department of Mathematics and Statistics, La Trobe University, Melbourne, VIC, 3086, Australia\\
             Tel.: +61394792609\\
              \email{a.olenko@latrobe.edu.au}          
}

\date{Received: date / Accepted: date}

\maketitle
\begin{abstract}
The paper starts by giving a motivation for this research and justifying the considered stochastic diffusion models for cosmic microwave background radiation studies. Then it derives the exact solution in terms of a series expansion to a hyperbolic diffusion equation on the unit sphere.
The Cauchy problem with random initial conditions is studied. All assumptions are stated in terms of the angular power spectrum of the initial conditions.
An approximation to the solution is given and analysed by finitely truncating the
series expansion. The upper bounds for the convergence rates of the approximation errors are derived. Smoothness properties of the solution and its approximation are investigated.  It is demonstrated that the sample H\"older continuity of these spherical fields is related to the decay of the angular power spectrum.
 Numerical studies of approximations to the solution and applications to cosmic microwave background data are presented to illustrate the theoretical results.
\end{abstract}

\section{Introduction}

The linear telegraph equation was introduced in the 1880s in Heaviside's model of transmission lines (see, e.g. \cite{Berg}). Since then, the same linear partial differential equation has arisen in several quite different contexts. 
In the 1950s, Cattaneo \cite{Cattaneo} introduced the hyperbolic heat equation that has a bounded speed of propagation of temperature disturbances, unlike the classical parabolic heat equation that has an unbounded propagation speed.  A bounded speed of propagation {also prohibits the generation of new cosmic structures that are correlated over space-like separated regions}. The large-scale coherent structures that are observed in the cosmic microwave background, are understood to be remnants of acoustic waves in the plasma universe, seeded by a very short inflationary period of superluminal expansion (e.g. \cite{Dodelson}, \cite{Weinberg}). Restricting the post-inflation propagation of disturbances to sub-luminal speeds is guaranteed by choosing the parameter  $c$ in the hyperbolic heat equation to be less than or equal to the speed of light. This has some rationale in relativistic geometry, and it maintains the second law of thermodynamics \cite{Ali}.

In Section 2, we add another application of the same equation. When $c$ is chosen to be the speed of light, the hyperbolic heat equation is indeed equivalent, by choosing an appropriate material coordinate system with conformal time coordinate, to the general relativistically covariant scalar Klein-Gordon equation, minimally coupled to an expanding space-time. In that coordinate system, the material radius of the expanding universe is constant. 

Our focus in later sections is on the hyperbolic diffusion of random disturbances on the  {surface or interior} of a sphere. In all of the aforementioned applications, this may be viewed as a canonical initial value problem on a compact manifold, as expressed in the classic texts such as \cite{Carslaw}. However, unlike the classic texts, we are  interested more in random initial conditions that include structures that cannot be causally connected. The evolution of fields and their correlations, under speed-limited diffusion, is of primary interest. In the case of the cosmic microwave temperature, typically with relative fluctuations of the order of $10^{-4}$, the currently available data are indeed represented on a spherical surface, with little reference to a radial coordinate, see \cite{PLANCK2}, \cite{PLANCK1},\cite{MarinucciPeccati11}. 

\begin{figure}[th]\vspace{-4mm}
	\begin{minipage}{0.5\textwidth}
		\includegraphics[trim = 3mm 43mm 32mm 38mm, clip, width=1\textwidth]{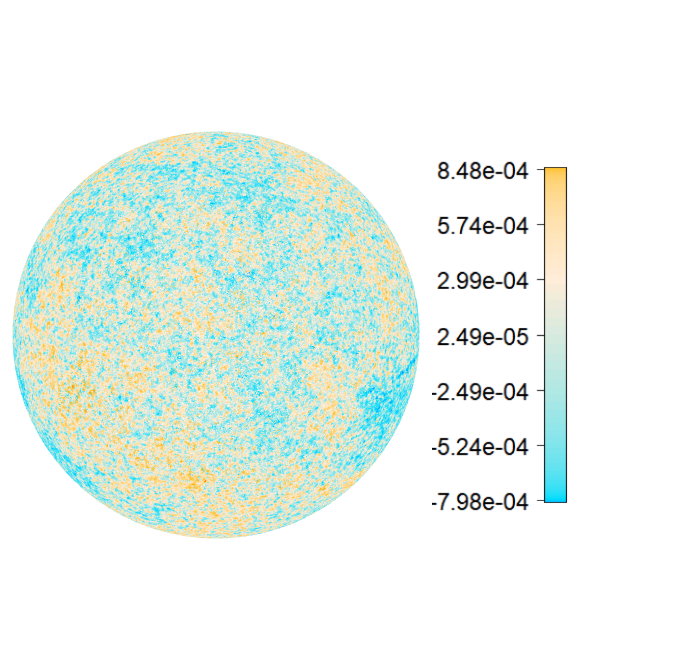}\\
		\caption{SMICA CMB intensity map at $\quad$ Nside = 1024 with 10 arcmin resolution}\label{cmbmap}
	\end{minipage}\hspace{1mm}
	\begin{minipage}{0.49\textwidth}
		\includegraphics[trim = 4mm 10mm 10mm 15mm,clip, width=1\textwidth,height=0.8\textwidth]{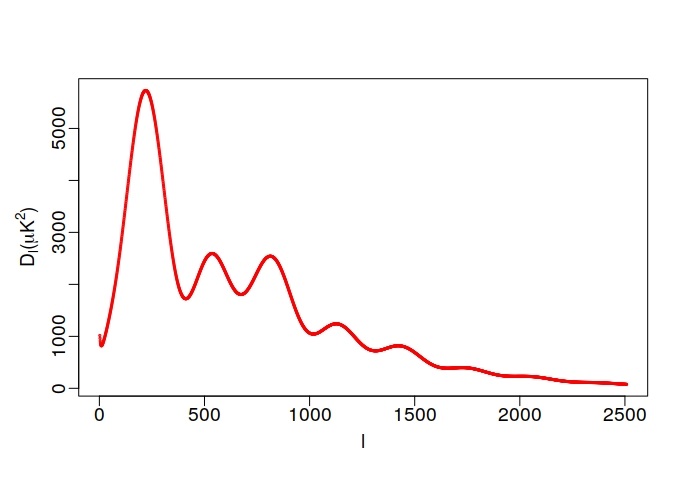}\\[-2mm]
		\caption{Best-fit LCDM CMB power spectra from the baseline Planck}\label{ang_spec}
	\end{minipage}
\end{figure}

Recent years have witnessed an enormous amount of attention, in the
astrophysical and cosmological literature, on investigating
spherical random fields. The empirical motivation for these studies comes from the current cosmological research. The NASA satellite mission WMAP and the ESA mission Planck, see \cite{PLANCK2}, \cite{PLANCK1},  probe Cosmic Microwave Background radiation
(CMB) to an unprecedented accuracy. Figure~\ref{cmbmap} shows measurements of the CMB temperature intensity from Planck 2015 results used as an illustration in this paper.  Figure~\ref{ang_spec} plots the corresponding best-fit scaled angular power spectrum.
CMB can be viewed as a signature of the
distribution of matter and radiation in the very early universe, and as such
it is expected to yield very tight constraints on physical models for the
Big Bang  { and subsequent phase transitions and nucleosynthesis}. For the density fluctuations of this field, the highly popular
inflationary scenario predicts a Gaussian distribution, whereas alternative
cosmological theories, such as topological defects or non-standard
inflationary models, predict otherwise.  

The seminal work of Applegate~et.~al. \cite{Applegate} showed that neutron-rich regions due to the longer mean free path of neutrons compared to protons, could strongly influence the distribution of helium and  deuterium during nucleosynthesis. When the temperature cooled  to allow weak nuclear re-combination and freeze-out, ($\frac 32 kT< 800$ keV), the neutron diffusion length was around 0.08 pc or 0.3 light-years \cite{Kurki}. Some basic considerations on diffusion length are given in Appendix A. After allowing for  neutron back-diffusion to proton-rich regions where neutrons have been further depleted by fusion \cite{Terasawa}, the observed distribution of elements and temperature variations  constrains the parameters of  heterogeneous models \cite{Barrow}, \cite{Iocco}.

Fluctuations in CMB observations may also have a
non-physical origin, i.e. they might be generated by systematic errors in
the CMB map, such as noise which has not been properly removed,
contamination from the galaxy or distortions in the optics of the telescope.
A proper understanding of the density distributions of fluctuations is also
instrumental for correct inference on the physical constants which can be
estimated from CMB radiation.

From a mathematical point of view, properties of random fields defined by SPDE on Euclidean spaces is a well studied area, see, for example, \cite{Anh0} and the references therein. However the known results in the literature are not directly translatable to manifolds. Therefore, the problems of approximating and studying sample regularity of random fields on the sphere have attracted much recent attention, see, for example,  \cite{Anh}, \cite{Lang}, \cite{MarinucciPeccati11}, \cite{Xiao} and the references therein. It was shown that the convergence rate of approximation schemes based on truncated series expansions of such fields is often related to the decay of the angular power spectrum. Another line of investigations started in the paper \cite{Kozachenko}. It studied approximations of Gaussian isotropic random fields on the sphere, but used different models and types of convergence. In this research we continue these investigations for spherical random fields generated by stochastic hyperbolic diffusion equations.

The discussion in the paper and the obtained results give some indication that the considered stochastic hyperbolic diffusion equation on the sphere can be used to capture important statistical and spectral properties of the CMB. The current research uses a different approach and a stochastic model comparing to \cite{Anh},\cite{Kozachenko}, \cite{Lang},  and \cite{Xiao}. As the approach yields the explicit solution in terms of series of elementary functions it could be very useful for various qualitative and numerical studies. The numerical studies suggest that the proposed approximations to solutions have an optimal order of convergence.

The paper is organised as follows. Section~\ref{sec2} and Appendix A provide physical motivation and detailed justification for stochastic models studied in the paper. Basic results and definitions about spherical isotropic random fields and their spectral and covariance representations are given in Section~\ref{sec3}. Section~\ref{sec4} first derives the solution of the non-random hyperbolic diffusion equation on the sphere.  Then, it gives the main results about solutions of the  hyperbolic diffusion equation with random initial conditions. The convergence rates of truncated solutions to the exact solution are derived in Section~\ref{secapprox}. Smoothness properties of the solution and its approximation are studied in the next section.
Section~\ref{sec6} and Appendix C present numerical studies and applications to CMB data. Short conclusions and future directions are discussed in Section~\ref{sec8}. All  proofs are given in Appendix B.

All numerical computations and simulations in this paper were performed using the software R version 3.5.3 and Python version 3.6.7. The results were derived using the HEALPix representation of the CMB data, see \cite{gor} and \mbox{\url{http://healpix.sourceforge.net}}. In particular, the R package rcosmo \cite{Fryer 2}, \cite{Fryer} was used for accessing the CMB data, computations and visualisations of the obtained results. The Python package healpy  was used for fast spherical harmonics transformations of CMB maps.  The R and Python code used for numerical examples in Section~\ref{sec6} are freely available in the folder "Research materials" from the  website \mbox{\url{https://sites.google.com/site/olenkoandriy/}}

\section{Derivation of hyperbolic diffusion equations}\label{sec2}

This section presents physical motivation and detailed justification for mathematical and stochastic models considered in the following sections. It also discusses and provides values of the model parameters for numerical studies in Section~\ref{sec6}.

Conduction of heat results from energy transfer during collisions of constituent particles or, in rigid lattices, the transfer of photons of vibrational energy from one lattice site to another. These transfer processes have a typical time delay $\tau_0$, leading to a relaxation time $\tau_0$ in a continuum dynamical model. For example, particles in a Boltzmann gas of hard spheres of radius $r_0$ have mean kinetic energy $\frac 12 mv^2=\frac 3 2 k_Bu$ and mean free path is $\ell=k_Bu/4\sqrt 2\pi r_0^2p$ where $k_B$ is Boltzmann's constant, $u({\bf x},t)$ is absolute temperature, $v({\bf x},t)^2$ is mean squared velocity and $p({\bf x},t)$ is pressure \cite{Thompson}. This means that the average time before a given particle collides is $\tau_0=\ell/v$. Before this time, the temperature gradient has no effect on the energy flux. This would lead to a delayed PDE but as demonstrated by Cattaneo \cite{Cattaneo}, it is convenient to approximate a delayed differential equation by a higher-order differential equation that is local in time. From the lag between instantaneous heat flux density ${\bf q}({\bf x},t)$ and the gradient of instantaneous temperature  $u({\bf x},t)$, there follows a higher-order correction to Fourier's law simply by Taylor expansion:
\begin{equation}
 {\bf q}({\bf x},t+\tau_0)\approx\tau_0\frac{\partial \bf q({\bf x},t)}{\partial t}+{\bf q({\bf x},t)}=-k\nabla u({\bf x},t),
 \end{equation}
where ${\bf q}({\bf x},t), {\bf x} \in \mathbb R^3, t>0,$ is a vector field, $u({\bf x},t), {\bf x} \in \mathbb R^3, t>0,$ is a scalar field, $\tau_0$ and $k$ are constants.

This is a two-parameter generalisation of Fourier's law, wherein $k$ is the thermal conductivity.
Then by the local conservation of heat energy, $\rho C\frac{\partial u ({\bf x},t)}{\partial t}=-\nabla\cdot {\bf q}({\bf x},t)$, where $\rho$ is density of the medium and $C$ is specific heat capacity. This leads directly to Cattaneo's hyperbolic heat/diffusion equation,
\begin{equation}
\label{HHE}
\frac{1}{c^2}\frac{\partial^2 u({\bf x},t)}{\partial t^2}+\frac{1}{D}\frac{\partial u({\bf x},t)}{\partial t}=\nabla^2 u({\bf x},t) ,
\end{equation}
where $D=k/C\rho$ and $c=\sqrt{D/\tau_0},$ which is the least upper bound for the speed of propagation. This can easily be seen by constructing attenuated travelling sinusoidal wave solutions by separation of variables.

The simplest relativistic extension of the non-relativistic Schr\"odinger  matter wave equation for a particle of mass $m$ is the Klein-Gordon equation in flat Minkowski space,
 $$\frac{1}{c^2}\frac{\partial ^2\varphi({\bf x},t) }{\partial t^2}-\nabla^2\varphi({\bf x},t) +\frac{m^2c^2}{\hbar^2}\varphi({\bf x},t) =0,$$
 which describes a spin-zero matter field \cite{Wigner} such as a Higgs scalar Boson \cite{Higgs}. In standard physics notation, $\hbar=h/2\pi$, where $h$ is Planck's constant.
 The Klein-Gordon equation agrees with the Schr\"odinger equation at low energies  (e.g. \cite{Bjorken}). The scalar field $\varphi=\varphi({\bf x},t), {\bf x} \in \mathbb R^3, t>0,$ has an interpretation in quantum field theory after second quantisation when it is regarded as an operator rather than a classical function \cite{Schweber}.
 
 Now consider a scalar field minimally coupled to a spatially flat de Sitter universe, with expansion factor $a(t)$, depending on cosmic time $t$. The space-time metric is
 \begin{eqnarray*}
 	ds^2&=&g_{\mu\nu}dx^\mu dx^\nu= c^2dt^2-a^2(t)\sum_{i=1}^3 dx^idx^i \\
 	&=&a^2(\eta)\left(c^2d\eta^2-\sum_{i=1}^3dx^idx^i\right).
 \end{eqnarray*}
 Here, $x^i$ are material coordinates. For example, these would remain constant at the centre of mass of a typically moving galaxy, where $t$ is the proper time. $\eta$ is the conformal time coordinate, $\eta=\int [1/a(t)]dt $, the adoption of which renders the metric to be conformal to the Minkowski metric. In this coordinate system $g_{\mu\nu}$
 is a diagonal matrix with entries $g_{00}=c^2a^2$ and $g_{jj}=-a^2$ for $j=1,2,3.$ 
When adapted to the more general space-time, the scalar covariant Klein-Gordon wave equation is (e.g.  \cite {Birrell}) 
\begin{equation}
g^{\mu\nu}\nabla_\mu\nabla_\nu\varphi({\bf x},t)+\left(\frac{m^2c^2}{\hbar^2}+\xi R\right)\varphi({\bf x},t)=0,
\label{KG}
\end{equation}
where $\nabla_\mu$ is the covariant derivative with respect to $x^\mu$ and $g^{\mu\nu}$ is the matrix inverse of $g_{\mu\nu}$.

We consider the case of negligible quantum particle mass $m$, and zero coupling ($\xi=0$) to the Ricci scalar $R$, in (\ref{KG}). This case is still called "minimal coupling" because the space-time metric influences the covariant derivative and the Laplace-Beltrami operator. In  \cite{PBPZ}, it was convenient to analyse the unattenuated field $a\varphi({\bf x},t).$ In the current context, we retain as the field of interest, $\varphi({\bf x},t)$ which is attenuated as the universe expands.

It is well known (e.g. \cite{Birrell}) that the Laplace-Beltrami operator, which is the second-order operator acting in the first term of (\ref{KG}), can be expressed more conveniently in terms of the determinant of the metric tensor, $\det g=\det [g_{\mu\nu}]$. Then (\ref{KG}) is equivalent to 
\begin{equation}
|\det g|^{-1/2}\frac{\partial}{\partial x^\mu}\left[|\det g|^{1/2}g^{\mu\nu}\frac{\partial \varphi({\bf x},t)}{\partial x^\nu}\right]+\left(\frac{m^2c^2}{\hbar^2}+\xi R\right)\varphi({\bf x},t)=0.
\label{KG2}
\end{equation}
For the  expanding spherically symmetric universe with $\det g=-c^2a^8$, a direct calculation gives
\begin{equation}
\label{KGPDE}
\frac{1}{c^2}\left[\frac{2a'(\eta)}{a}\frac{\partial\varphi({\bf x},t)}{\partial\eta}+\frac{\partial^2\varphi({\bf x},t)}{\partial\eta^2}\right]-\sum_{i=1}^3\frac{\partial^2\varphi({\bf x},t)}{\partial x^i\partial x^i}=0.
\end{equation}
Near some time $t=t_1$ in the past, the expansion factor may be approximated by a linear function $a=a_1+c_1(t-t_1)/R_0$, where $R_0$ is the current radius and $c_1$ is the speed of expansion at time $t_1$. In that case,
$$\eta=\eta_1+\frac{R_0}{c_1}\ln\left(1+\frac{c_1[t-t_1]}{a_1R_0}\right),$$
$$a=a_1[1+e^{c_1(\eta-\eta_1)/R_0}],$$ and $a'(\eta)/a=\frac{c_1}{2R_0}$ at $t=t_1$. The covariant Klein-Gordon equation is approximated by the relativistic diffusion equation, which is the same as  the hyperbolic diffusion equation (\ref{HHE}) with  $D=c^2R_0/c_1$.
Up to the present time, CMB data is represented on a 2-sphere, with little reference to radial variation. This prompts one to take a volume average along the radial direction in a spherical sector with fixed small differential increment of solid angle, throughout (\ref{KGPDE}). Defining
	\[\label{radial}
	\bar\varphi(R,\theta,\phi,\eta)=\frac{3}{R^3}\int_0^R r^2 \varphi(r,\theta,\phi,\eta) dr,
	\] 
	where $(r,\theta,\phi)$ are material polar coordinates, (\ref{KGPDE}) implies 
\[ \frac{1}{c^2}\left[ \frac{2a'(\eta)}{a}\frac{\partial}{\partial\eta}+\frac{\partial^2}{\partial\eta^2}\right] \bar\varphi(R,\theta,\phi,\eta) -\left\{\frac{1}{R^2 \sin (\theta)}\frac{\partial}{\partial \theta}\left[ \sin (\theta) \frac{\partial }{\partial \theta}\right]\right.\] 
		\[\left.
	+\frac{1}{R^2\sin^2(\theta)}\frac{\partial^2}{\partial\phi^2}\right\}\bar\varphi(R,\theta,\phi,\eta) = \frac{3}{R}\frac{\partial\varphi(R,\theta,\phi,\eta)}{\partial R}\]
		\[+\frac{2}{R^3}\int_0^R\left(\frac{1}{\sin(\theta)}\frac{\partial}{\partial\theta}\left[\sin(\theta)\frac{\partial\bar\varphi(r,\theta,\phi,\eta)}{\partial\theta}\right]+\frac{1}{\sin^2(\theta)}\frac{\partial^2\bar\varphi(r,\theta,\phi,\eta)}{\partial\phi^2}\right)dr \]
	\[- \lim_{r\to  0}\left(\frac rR\right)^3\left\{\frac{1}{r^2\sin(\theta)}\frac{\partial}{\partial\theta}\left[\sin(\theta)\frac{\partial\bar\varphi(r,\theta,\phi,\eta)}{\partial\theta}\right]+\frac{1}{r^2\sin^2(\theta)}\frac{\partial^2\bar\varphi(r,\theta,\phi,\eta)}{\partial\phi^2} \right\}.
	\]
	The two limits on the right hand side sum to
		$-\lim_{r\to 0}\left(\frac rR\right)^3\frac{1}{r^2}\Delta_{ \mathbb S^2}\varphi,$ where $r^{-2}\Delta_{ \mathbb S^2}\varphi $ is the surface Laplacian on a sphere of vanishingly small radius. Since the Laplacian is not expected to be more singular than $\mathcal O(r^{-2})$, this limit is zero. The free boundary condition of zero mass-energy-flux across the expanding maximal radius $a(\eta)R$, reduces in material coordinates to $\frac{\partial\varphi}{\partial r}=0$ at $r=R$. This annuls the first term on the right hand side of the above. The remaining integral term is $2/3$ times the sector radially doubly averaged divergence of $\nabla\varphi$. The sector average is weighted towards values at large $r$. Double averages are weighted even more towards values at large $r$. Doubly averaged fluctuations in $\varphi$ are expected to be small compared to singly averaged fluctuations. As an indicative study of a scalar field, we set the right hand side of~(\ref{radial}) to zero. Then near $\eta=\eta_1$, the equation for $\bar\varphi$  reduces to 
\begin{equation}
\label{CCHDE}
\frac{1}{c^2}\frac{\partial^2\bar\varphi(R,\theta,\phi,\eta)}{\partial\eta^2}+\frac{c_1}{c^2R_0}\frac{\partial\bar\varphi(R,\theta,\phi,\eta)}{\partial\eta}=\frac{1}{R_0^2}\Delta_{\mathbb S^2}\bar\varphi (R,\theta,\phi,\eta).
			\end{equation}  
			Note that $R_0^{-2}\Delta_{\mathbb S^2}$ is the Laplace-Beltrami operator on a spherical surface of radius $R_0$. In the material coordinate system, the material radius is constant $R_0$ at all times. There is a unique rescaling of variables that further reduces  (\ref{CCHDE}) to normalised form,
\[
			\frac{\partial^2\bar\varphi(R,\theta,\phi,\hat\eta)}{\partial\hat\eta^2}+\frac{\partial\bar\varphi(R,\theta,\phi,\hat\eta)}{\partial\hat\eta}=\frac{1}{\hat R^2}\Delta_{\mathbb S^2}\bar\varphi(R,\theta,\phi,\hat\eta),\]
			\[	
			\hat\eta=\eta/\eta_s=\eta c^2/D=\eta c_1/R_0~;~~\hat R=R_0/\ell_s=R_0c/D=c_1/c.	
\]

				The single remaining parameter $\hat R$ involves the expansion rate at time $t_1$ which is related to the Hubble parameter. In particular, we are interested in a time $t_1$ when the cosmic electromagnetic radiation began to propagate through a transparent recombined atomic medium, whose properties may still be inferred.
				
				The cosmic background electromagnetic radiation became free  for unimpaired propagation into the distant future, around the narrow band of time of atomic recombination. Seemingly by coincidence, this narrow period was close to the time of equality of matter density and radiation density expressed in the same units  (e.g. \cite{Hir}).  Henceforth,  subscript zero will denote a parameter value measured at the current time in history whereas a subscript $e$ will denote its value at the time of equality. At the present time, the Hubble constant is measured to be $H_0=\dot a(t)/a(t)=0.7$ km/s/Mpc and the Hubble time is $1/H_0=1.4\times 10^{10}$ yr. From the Friedman equation, the Hubble `constant' $H$ actually depends on time according to 
$$H^2=H_0^2(\Omega_\Lambda+\Omega_ma(t)^{-3}+\Omega_{\gamma\nu}a(t)^{-4}+\Omega_\kappa a(t)^{-2}).$$ The various components $\Omega_J$ are fractions of current total energy that are dark energy $\Omega_\Lambda$,  mass  density $\Omega_m$, radiation energy density $\Omega_{\gamma\nu}$ and curvature energy density $\Omega_\kappa$. Within experimental error, the universe is spatially flat (e.g. NASA WMAP website \cite{NASA}), so  the total mass-energy is the critical energy for the universe to be spatially flat. The present estimates, with $a(t)=1$, have dark energy $\Omega_\Lambda =0.714$, and matter $\Omega_m$=0.286 consisting of dark matter (0.240) and baryons (0.046). The component $\Omega_\kappa$ due to spatial curvature is measured to be zero and the component due to combined photon and neutrino radiation ($\Omega_{\gamma\nu}$) is negligible at the current time. However at other times, the components of mass-energy, relative to the current critical mass-energy, are multiplied by various powers of $a(t)$. For example around the age of 370,000 years after the big bang, within the short period of atomic recombination, the radiation and matter components $\Omega_m a(t)^{-3}$ and $\Omega_{\gamma\nu}a(t)^{-4}$ were equal and the other components were negligible. This occurs at expansion factor   $a_e=\Omega_{\gamma\nu} / \Omega_m =4.15\times 10^{-5}$. Consequently, \\
				\[
				H_e=(2\Omega_m)^{1/2}a_e^{-3/2}H_0,\quad c_e=\dot a_e\, a_eR_0,\quad
				D=\frac{c^2R_e}{c_e}=\frac{c^2}{H_ea_e}.
				\]
				The conformal time scale appearing above is $\eta_s=D/c^2=R_e/c_e=1/H_e$, which is exactly the Hubble time at $t=t_e$. This is approximately $9\times 10^{-8}/H_0$, where $1/H_0$ is the current Hubble time. The material length scale evaluates to $\ell_s=\frac{4.15^{3/2}}{2\times 5.72^{1/2}}\times 10^{-7}c/H_0.$
				This is approximately 0.02 times the radius of the universe at recombination time.
\section{Isotropic random fields}\label{sec3}
This section introduces basic notations and background by reviewing some results in the theory of spherical random fields from the monograph 
\cite{Yadrenko} (see, also \cite{Ivanov}, \cite{Lang}, \cite{Leonenko99}, \cite{MarinucciPeccati11}).

We will use the symbol $C$ to denote constants which are not important for our exposition. Moreover, the same symbol may be used for different constants appearing in the same proof.

Consider a sphere in the three-dimensional Euclidean space%
\[
\mathbb{S}^2=\left\{ \mathbf{x}\in \mathbb{R}^{3}:\Vert \mathbf{x}\Vert =1\right\} \subset \mathbb{R}%
^{3} \nonumber
\]
with the Lebesgue measure (the area element on the sphere)%
\[
\widetilde{\sigma }(d\mathbf{x})=\sigma (d\theta, d\varphi )=\sin \theta d\theta
d\varphi , \quad (\theta ,\varphi )\in [0,\pi)\times[0,2\pi). 
\]%
A spherical random field on a complete probability space $(\Omega ,\mathcal{F%
},\mathbf{P})$, denoted by%
\[
T=\left\{ T(\theta ,\varphi )=T_{\omega }(\theta ,\varphi ):0\leq \theta
< \pi ,\quad 0\leq \varphi < 2\pi ,\ \omega \in \Omega \right\} , 
\]%
in the spherical coordinate system, or $\widetilde{T}=\{\widetilde{T}(\mathbf{x})$ $,$ $\mathbf{x}\in \mathbb{S}^2\}$ in the Cartesian coordinates,  is a stochastic function
defined on the sphere $\mathbb{S}^2.$ 

We consider a real-valued second-order spherical random
field $T$ that is continuous in the
mean-square sense. Note that \cite{MarinucciPeccati13} proved that the
covariance function of a measurable finite-variance isotropic random field
on the sphere is necessarily everywhere continuous.

Under these conditions, the field $T$ can be expanded in the mean-square
sense as a Laplace series (see, \cite{Yadrenko}, p. 73, \cite%
{Leonenko99}, p. 33, or \cite{MarinucciPeccati13}, p.123): 
\begin{equation}  \label{2.1}
T(\theta ,\varphi )=\sum_{l=0}^{\infty }\sum_{m=-l}^{l}a_{lm}Y_{lm}(\theta
,\varphi ),
\end{equation}
where $\{Y_{lm}(\theta ,\varphi )\}$ represents the complex spherical harmonics.
The spectral representation~(\ref{2.1}) can be viewed as a Karhunen-Lo\`{e}ve
expansion, which converges in the Hilbert space $L_{2}(\Omega \times
\mathbb{S}^2,\sin \theta d\theta d\varphi )$, that is,%
\[
\lim_{L\rightarrow \infty }\mathbf{E}\left( \int\limits_{\mathbb{S}^2}\left(
T(\theta ,\varphi )-\sum_{l=0}^{L}\sum_{m=-l}^{l}Y_{lm}(\theta ,\varphi
)a_{lm}\right) ^{2}\sin \theta d\theta d\varphi \right) =0. 
\]%
According to the Peter-Weyl theorem (see \cite{MarinucciPeccati13}, p.69), the
expansion (\ref{2.1}) also converges in the Hilbert space $L_{2}(\Omega ),$
for every $\mathbf{x}\in \mathbb{S}^2$, that is, for each $\mathbf{x}\in \mathbb{S}^2,$ 
\[
\lim_{L\rightarrow \infty }\mathbf{E}\left( \widetilde{T}(\mathbf{x})-\sum_{l=0}^{L}%
\sum_{m=-l}^{l}\tilde{Y}_{lm}(\mathbf{x})a_{lm}\right) ^{2}=0, 
\]
where $\{\tilde{Y}_{lm}(\mathbf{x})\}$ represents the complex spherical harmonics of the Cartesian variable $\mathbf{x}.$

Recall that for $-l\leq m\leq l$ it holds
\[
\tilde{Y}_{lm}(\mathbf{x})=Y_{lm}(\theta ,\varphi )=d_{lm}\exp(im\varphi
)P_{l}^{m}(\cos\theta),
\]
\[d_{lm}=(-1)^{m}\left[\frac{(2l+1)(l-m)!}{4\pi(l+m)!}\right] ^{1/2},  \label{2.2}\nonumber
\]
where $P_{l}^{m}(\cdot)$ denotes the associated Legendre polynomial with the indices $l$ and $m,$
and $P_l(\cdot)$ is the $l$-th Legendre polynomial, i.e. 
\begin{equation}
P_{l}^{m}(x)=(-1)^{m}(1-x^{2})^{m/2}\frac{d^{m}}{dx^{m}}P_{l}(x), \quad
P_{l}(x)=\frac{1}{2^{l}l!}\frac{d^{l}}{dx^{l}}(x^{2}-1)^{l}.  \label{2.3}
\end{equation}%
The spherical harmonics have the following properties%
\[
\int_{0}^{\pi }\int_{0}^{2\pi }Y_{lm}^{\ast }(\theta ,\varphi )Y_{l^{\prime
	}m^{\prime }}(\theta ,\varphi )\sin \theta d\varphi d\theta  =\delta
_{l}^{l^{\prime }}\delta _{m}^{m^{\prime }},  \]
\begin{equation}\label{Ysym}
Y_{lm}^{\ast }(\theta ,\varphi ) =(-1)^{m}Y_{l(-m)}(\theta ,\varphi ),
\end{equation}
\[Y_{lm}(\pi -\theta ,\varphi +\pi ) = (-1)^{l}Y_{lm}(\theta ,\varphi ), \]
\begin{equation}\label{Yl0}
\tilde{Y}_{l0}(\mathbf{0})=\sqrt{\frac{2l+1}{4\pi}}P_{l}(1)=\sqrt{\frac{2l+1}{4\pi}},
\end{equation}
where $\delta _{l}^{l^{\prime }}$ is the Kronecker delta function,  the symbol * means the complex conjugation and $\mathbf{0}$ corresponds to $\varphi=\theta=0.$ The random
coefficients $a_{lm}$ in the Laplace series~(\ref{2.1}) can be obtained through
inversion arguments in the form of mean-square stochastic integrals%
\begin{equation}
a_{lm}=\int_{0}^{\pi }\int_{0}^{2\pi }T(\theta ,\varphi)Y_{lm}^{\ast
}(\theta ,\varphi )\sin \theta d\theta d\varphi .  \label{2.5}
\end{equation}%

As $T$ is real-valued, then, by the property (\ref{Ysym}),  it holds
\begin{equation}
a_{lm} = (-1)^m a_{l\,-m},\quad l \ge 1,\ -l\leq m\leq l.\label{almreal}
\end{equation}

The field $\widetilde{T}(\mathbf{x})$ is called isotropic (in
the weak sense) on the sphere $\mathbb{S}^2$ if $\mathrm{E}\widetilde{T}%
(\mathbf{x})^{2}<\infty$ and its first and second-order moments are invariant with
respect to the group $SO(3)$ of rotations in $\mathbb{R}^{3},$ i.e. 
\[
\mathbf{E}\widetilde{T}(\mathbf{x})= \mathbf{E}\widetilde{T}(g\mathbf{x}),\quad  \mathbf{E}%
\widetilde{T}(\mathbf{x})\widetilde{T}(\mathbf{y})= \mathbf{E}\widetilde{T}(g\mathbf{x})\widetilde{T}%
(g\mathbf{y}), 
\]%
for every $g\in SO(3)$ and  $\mathbf{x}, \mathbf{y} \in \mathbb{S}^2.$  This is
equivalent to saying that the mean $ \mathbf{E}T(\theta ,\varphi )=c=constant$
(without loss of generality we assume that $c=0),$ and that the covariance function $ \mathbf{E}T(\theta
,\varphi )T(\theta ^{\prime },\varphi ^{\prime })$ depends only on the
angular distance $\Theta =\Theta _{PQ}$ between the points $P=(\theta
,\varphi )$ and $Q=(\theta ^{\prime },\varphi ^{\prime })$ on $\mathbb{S}^2.$ 

The
field is isotropic if and only if 
\[
\mathbf{E}a_{lm}a_{l^{\prime }m^{\prime }}^{\ast }=\delta _{l}^{l^{\prime
}}\delta _{m}^{m^{\prime }}C_{l},\quad -l\leq m\leq l,\quad -l^{\prime }\leq
m^{\prime }\leq l^{\prime }. 
\]%
Thus, $ \mathbf{E}|a_{lm}|^{2}=C_{l},$ $m=0,\pm 1,...,\pm l.$ The series $\left\{ C_{1},C_{2},...,C_{l},...\right\}$ is called the
angular power spectrum of the isotropic random field $T(\theta ,\varphi ).$

From (\ref{2.1}) and (\ref{2.5}) we deduce that the covariance function of an isotropic random fields has the  following representation
\[
\Gamma(\cos\Theta )= \mathbf{E}T(\theta ,\varphi )T(\theta ^{\prime },\varphi
^{\prime})=\frac{1}{4\pi }\sum_{l=0}^{\infty}(2l+1)C_{l}P_{l}(\cos\Theta),
\label{2.8}\]
where%
\begin{equation}
\sum_{l=0}^{\infty }(2l+1)C_{l}<\infty.  \label{SER}
\end{equation}%
If $T(\theta ,\varphi )$ is an isotropic Gaussian
field, then the coefficients $a_{lm},$ $m=-l,\dots ,l,$ $l\geq 1,$ are
complex-valued independent Gaussian random variables if $m\not=-m^{\prime },$
with 
\begin{equation}\label{alm}
\mathbf{E}a_{lm}=0,\quad  \mathbf{E}a_{lm}a_{l^{\prime }m^{\prime }}^{\ast
}=\delta _{m}^{m^{\prime }}\delta _{l}^{l^{\prime }}C_{l}, 
\end{equation}%
if $C_{l}>0$, or degenerate to zero if $C_{l}=0.$ 

\section{Solution for stochastic spherical hyperbolic diffusion}\label{sec4}

First this section derives solutions of non-random hyperbolic diffusion equations. Then the obtained results are used in to obtain solutions of diffusion equations with random initial conditions. 

Consider the following hyperbolic diffusion equation, also known as the
telegraph equation (see \cite{Kolesnik}) or relativistic diffusion equation on sphere 
\begin{equation}
\frac{1}{c^{2}}\frac{\partial ^{2}\tilde{p}(\mathbf{x},t)}{\partial t^{2}}+%
\frac{1}{D}\frac{\partial \tilde{p}(\mathbf{x},t)}{\partial t}=k^{2}\Delta
_{\mathbb{S}^2}\;\tilde{p}(\mathbf{x},t),  \quad t\geq 0,  \label{teleq}
\end{equation}%
with the initial conditions 
\begin{equation}
\tilde{p}(\mathbf{x},t)|_{t=0}=\delta (\mathbf{x}),\qquad \left. \frac{%
	\partial \tilde{p}(\mathbf{x},t)}{\partial t}\right\vert _{t=0}=0,
\label{incond}
\end{equation}%
where $\mathbf{x}=(x_{1},x_{2},x_{3})\in \mathbb{S}^2,$ $c>0,\;D>0,$ and $k$ are some constants, and $\Delta _{\mathbb{S}^2}$ is the
Laplace operator on the sphere $\mathbb{S}^2$ and $\delta (\mathbf{x})$ is the
Dirac delta-function.

Note, that in the unit spherical coordinates equation (\ref%
{teleq}) takes the form 
\[
\frac{1}{c^{2}}\frac{\partial ^{2}p(\theta ,\varphi ,t)}{\partial t^{2}}+%
\frac{1}{D}\frac{\partial p(\theta ,\varphi ,t)}{\partial t}=k^{2}\Delta
_{(\theta ,\varphi )}\;p(\theta ,\varphi ,t),\]
\[
\theta \in \lbrack 0,\pi ),\;\varphi \in \lbrack 0,2\pi ),\;t>0, 
\]%
where 
\begin{equation}  \label{lapbel}
\Delta _{(\theta ,\varphi )}=\frac{1}{\sin \theta } \; \frac{\partial }{
	\partial \theta }\left( \sin {\theta }\;\frac{\partial }{\partial \theta }
\right) +\frac{1}{\sin ^{2}{\theta }}\;\frac{\partial ^{2}}{\partial
	\varphi^{2}}
\end{equation}
is the Laplace-Beltrami operator on the sphere.

It is known (see, i.e., \cite{MarinucciPeccati13}, p.72) that the eigenvalue
problem for Laplace operator on the sphere has the following exact solution 
\begin{equation}\label{sphharm}
\Delta _{\mathbb{S}^2}\;\tilde{Y}_{lm}(\mathbf{x})=-l(l+1)\tilde{Y}_{lm}(\mathbf{x}),\qquad
l=0,1,2,\dots ,\quad m=-l,\dots  ,l, 
\end{equation}%
where $\{\tilde{Y}_{lm}(\mathbf{x})\}$ is the system of spherical harmonics.
Therefore, it is natural to seek a solution of the problem (\ref%
{teleq})-(\ref{incond}) in the form of the series 
\begin{equation}
\tilde{p}(\mathbf{x},t)=\sum_{l=0}^{\infty }\sum_{m=-l}^{l}b_{lm}(t)\;\tilde{Y}_{lm}(%
\mathbf{x}),  \label{series}
\end{equation}%
where 
\begin{equation}
b_{lm}(t)=\int_{\mathbb{S}^2}\tilde{p}(\mathbf{x},t)\;\tilde{Y}_{lm}^{\ast }(\mathbf{x})\;%
\tilde{\sigma}(d\mathbf{x}),  \label{coeff}
\end{equation}%
$\tilde{\sigma}(d\mathbf{x})=\sin \theta d\theta d\varphi $.

The proof of the following result is given in Appendix B.

\begin{theorem}\label{Th1} The solution $\tilde{p}(\mathbf{x},t)$ to the hyperbolic diffusion point-source initial value problem {\rm (\ref{teleq})-(\ref{incond})}  is given by the series 
	\[  
 \tilde{p}(\mathbf{x},t) = \exp\left(-\frac{c^{2}t}{2D}\right)
	\sum_{l=0}^{\infty} Q_{l}(\mathbf{x})\Biggl( \biggl[ \cosh \left( tK_l\right)  + \frac{c^{2}}{2DK_l}\;\sinh
	\left( t K_l\right) \biggr]\]
	\[\times \mathbf{%
		1}_{\left\{ l\leq \frac{\sqrt{D^{2}k^{2}+c^{2}}-Dk}{2Dk}\right\} }  + \biggl[ \cos \left( tK_l'\right) + \frac{c^{2}}{2D K_l'}  \sin
	\left( t K_l'\right) \biggr] \mathbf{%
		1}_{\left\{ l> \frac{\sqrt{D^{2}k^{2}+c^{2}}-Dk}{2Dk}\right\} } \Biggr), %
	\]
	where $\mathbf{1}_{\{\cdot \}}$ denotes the binary indicator function,
	\[K_l=\sqrt{\frac{c^{4}}{4D^{2}}-c^{2}l(l+1)k^{2}},\quad K_l'=\sqrt{c^{2}l(l+1)k^{2}-\frac{c^{4}}{4D^{2}}},\]
\begin{equation}\label{Q}
Q_{l}(\mathbf{x})=\sum_{m=-l}^{l}\tilde{Y}_{lm}^{\ast }(\mathbf{0})\;\tilde{Y}_{lm}(\mathbf{x}). 
\end{equation}
\end{theorem}

Now we use the results of Theorem~\ref{Th1} to derive solutions of diffusion equations with random initial conditions.
The random field $u(\theta ,\varphi ,t)$ is defined by the following hyperbolic diffusion equation on the sphere 
\begin{equation}
\frac{1}{c^{2}}\frac{\partial ^{2}u(\theta ,\varphi ,t)}{\partial t^{2}}+%
\frac{1}{D}\frac{\partial u(\theta ,\varphi ,t)}{\partial t}=k^{2}\Delta
_{(\theta ,\varphi )}\;u(\theta ,\varphi ,t),  \label{TER1}
\end{equation}%
\[
\theta \in \lbrack 0,\pi ),\;\varphi \in \lbrack 0,2\pi ),\;t>0, 
\]%
where $\Delta _{(\theta ,\varphi )}$ is the Laplace-Beltrami operator on
the sphere given by (\ref{lapbel}). 

Now, the random initial conditions are determined by the Gaussian
iso\-tropic random field on the sphere 
\begin{equation}
u(\theta ,\varphi ,t)\big|_{t=0}=T(\theta ,\varphi )=\sum_{l=0}^{\infty
}\sum_{m=-l}^{l}a_{lm}Y_{lm}(\theta ,\varphi ),  \label{TER2}
\end{equation}

\begin{equation}
\left. \frac{\partial u(\theta ,\varphi ,t)}{\partial t}\right\vert _{t=0}=0,
\label{TER3}
\end{equation}%
where $a_{lm},$ $m=-l,\dots ,l,$ $l\geq 0,$ are complex-valued independent Gaussian
random variables satisfying (\ref{almreal}) and (\ref{alm}).

\begin{theorem}\label{Th2} If the angular power spectrum \{$C_{l},l=0,1,2,...\}$ of
	the random field $T(\theta ,\varphi )$ from the initial condition {\rm(\ref{TER2})} satisfies assumption {\rm(\ref{SER})}, then
	the random solution $u(\theta ,\varphi ,t)$ of the initial value problem {\rm(\ref{TER1})-(\ref{TER3})} is given by the convergent in the Hilbert space $L_{2}(\Omega \times \mathbb{S}^2,\sin\theta
		d\theta d\varphi)$ random series 
	\begin{equation}
	u(\theta ,\varphi ,t)= \exp\left(-\frac{c^{2}t}{2D}\right)
	\sum_{l=0}^{\infty}\sum_{m=-l}^{l}Y_{lm}(\theta ,\varphi )\xi _{lm}(t),
	\qquad t\geq 0,  \label{SOL}
	\end{equation}%
	where 
	\begin{equation}
	\xi _{lm}(t)= \sqrt{\frac{4\pi }{2l+1}}a_{lm}\tilde{Y}_{l0}^{\ast }(\mathbf{0}%
	)[A_{l}(t)+B_{l}(t)]  \label{xis}
	\end{equation}%
	are 
	stochastic processes with 
	\begin{equation}
	A_{l}(t) = \biggl[ \cosh \left( t K_l\right) + \frac{c^{2}}{2DK_l}\sinh \left( tK_l\right) \biggr] 
	\mathbf{1}_{\left\{ l\leq \frac{\sqrt{D^{2}k^{2}+c^{2}}-Dk}{2Dk}\right\} } \label{At}
	\end{equation}
	and 
	\begin{equation}
	B_{l}(t)  = \biggl[ \cos \left( t K_l'\right) + \frac{c^{2}}{2DK_l'}\sin \left( tK_l'\right) \biggr] 
	\mathbf{1}_{\left\{ l>\frac{\sqrt{D^{2}k^{2}+c^{2}}-Dk}{2Dk}\right\} }. \label{Bt}
	\end{equation}
	 Moreover, its covariance function is given by \[ \mathbf{Cov}(u(\theta ,\varphi ,t),u(\theta ^{\prime },\varphi ^{\prime },t^{\prime}))  = 
	 (4\pi)^{-1} \exp\left( -\frac{c^{2}}{2D}(t+t^{\prime })\right)\]
	 \begin{equation}\label{COV} 
	 \times\sum_{l=0}^{\infty}(2l+1)C_{l}
	 P_l(\cos\Theta) [A_{l}(t)A_{l}(t^{\prime})+B_{l}(t)B_{l}(t^{\prime })],
	 \end{equation}
\end{theorem}
The proof of Theorem~\ref{Th2} is given in Appendix B.

\section{Convergence study of approximate solutions}\label{secapprox}

The results in Section~\ref{sec4} provide a series representation of the random field $u(\theta ,\varphi ,t).$ To investigate contributions of different terms in the representation it is important to study their finite cumulative sums. In this section it is done by analysing truncated series expansions of the solution $u(\theta ,\varphi ,t)$  of the initial value problem {\rm(\ref{TER1})-(\ref{TER3})}. We demonstrate the role of the decay rate of the angular power spectrum. These results are also important for various numerical studies. In particular, they can be used to determine the  required number of terms in the truncated series to get a specified accuracy  of the approximate solutions.
 
The approximation $u_L(\theta ,\varphi ,t)$ of truncation degree $L\in\mathbb{N}$ to the solution $u(\theta ,\varphi ,t)$ in~(\ref{SOL}) is defined by
\[
u_L(\theta ,\varphi ,t) =  \exp\left(-\frac{c^{2}t}{2D}\right)
\sum_{l=0}^{L-1}\sum_{m=-l}^{l}Y_{lm}(\theta ,\varphi )\xi _{lm}(t),\]
for $\theta \in \lbrack 0,\pi ),$ $\varphi \in \lbrack 0,2\pi ),$ $t>0.$ 

The following theorem gives the convergence rate of the approximation $u_L(\theta ,\varphi ,t)$  to the solution $u(\theta ,\varphi ,t).$ The order of the convergence rate is determined by the high frequency magnitudes of the angular spectrum.  

\begin{theorem}\label{th3}
	Let $u(\theta ,\varphi ,t)$  be the solution to the initial value problem {\rm(\ref{TER1})-(\ref{TER3})} and  $u_L(\theta ,\varphi ,t)$ be the  approximation of truncation degree $L\in\mathbb{N}$ of $u(\theta ,\varphi ,t).$  
	
	Then, for $t>0$ the truncation error is bounded by
	\[
	\|u(\theta ,\varphi ,t)-u_L(\theta ,\varphi ,t)\|_{L_2(\Omega \times
		{\mathbb S}^{2})} \le  C\left(\sum_{l=L}^{\infty }(2l+1)C_{l}\right)^{1/2}.	
	\]
	
	Moreover, for $L>\frac{\sqrt{D^{2}k^{2}+c^{2}}-Dk}{2Dk}$ it holds	
	\begin{equation}\label{upb0}\|u(\theta ,\varphi ,t)-u_L(\theta ,\varphi ,t)\|_{L_2(\Omega \times
		{\mathbb S}^{2})}\le C\exp\left( -\frac{c^{2}t}{2D}\right)
	\left(\sum_{l=L}^{\infty}(2l+1)C_{l}\right)^{1/2},\end{equation}
	where the constant $C$ depends only on the parameters $c$, $D$ and $k.$
\end{theorem}

The proof of Theorem~\ref{th3} is given in Appendix B and does not depend on $\theta$ and $\varphi.$ Therefore, the statement also holds in $L_2(\Omega)$ norm  over the sphere ${\mathbb S}^{2}.$ Hence, we obtain the following results.
\begin{corollary} Uniformly over $\theta \in \lbrack 0,\pi ),$ $\varphi \in \lbrack 0,2\pi )$ the results of Theorem~{\rm \ref{th3}} are also valid for the mean squared truncation error
	\[\mathbf{MSE}(u(\theta ,\varphi ,t)-u_L(\theta ,\varphi ,t))=\mathbf{Var}\left(u(\theta ,\varphi ,t)-u_L(\theta ,\varphi ,t)\right)\]
	\[=\|u(\theta ,\varphi ,t)-u_L(\theta ,\varphi ,t)\|_{L_2(\Omega)}^2.\]
	
\end{corollary}

\begin{corollary}\label{cor2}
	Let the angular power spectrum  $\{C_{l},l=0,1,2,...\}$ of 
	the random field $T(\theta ,\varphi )$ from the initial condition
	{\rm(\ref{TER2})} decay algebraically with order $\alpha >2,$ that is, there exist constants $C >0$ and $l_0\in \mathbb{N}$ such that $C_{l}\le C\cdot l^{-\alpha}$ for all $l\ge l_0.$
	
	Then, the  approximation $u_L(\theta ,\varphi ,t)$ converges to   the solution $u(\theta ,\varphi ,t)$ of the initial value problem {\rm(\ref{TER1})-(\ref{TER3})} and
	\begin{enumerate}
		\item[{\rm(i)}]  for $L>\max(l_0,\frac{\sqrt{D^{2}k^{2}+c^{2}}-Dk}{2Dk})$ the truncation error is bounded by
		\[\|u(\theta ,\varphi ,t)-u_L(\theta ,\varphi ,t)\|_{L_2(\Omega \times
			{\mathbb S}^{2})}\le C\exp\left( -\frac{c^{2}t}{2D}\right)
		{L}^{-\frac{\alpha-2}{2}},\]
		\item[{\rm(ii)}] for any $\varepsilon>0$ it holds
		\[\mathbf{P}\Big(|u(\theta ,\varphi ,t)-u_L(\theta ,\varphi ,t)|\ge\varepsilon\Big)\le \frac{C\exp\left( -{c^{2}t}/D\right)}{{L}^{\alpha-2}\varepsilon^2},\]
		\item[{\rm(iii)}]  for all $\theta \in \lbrack 0,\pi ),$ $\varphi \in \lbrack 0,2\pi )$ and $t>0$ the truncation error is asymptotically almost surely bounded by
		\[|u(\theta ,\varphi ,t)-u_L(\theta ,\varphi ,t)|\le L^{-\beta} \quad \mathbf{P}-a.s.,\] where $\beta\in \left(0,\frac{\alpha -3}{2}\right)$ and $\alpha>3.$ 
	\end{enumerate} 
\end{corollary}
The proof of Corollary~\ref{cor2} is given in Appendix B.

\section{H\"older continuity of solutions and their truncated approximations} 
In this sections we investigate properties of the solution $u(\theta ,\varphi ,t)$  of the initial value problem {\rm(\ref{TER1})-(\ref{TER3})} and its approximation $u_L(\theta ,\varphi ,t).$ We show that they are H\"older continuous fields. It is demonstrated how the decay of the angular power spectrum is related to the H\"older continuity in mean square of the corresponding random field. 

It follows from \cite{MarinucciPeccati13} that the field $u(\theta ,\varphi ,t)$ is mean square continuous. However,  obtaining  sample H\"older continuity of this field requires stronger assumptions on the decay of the angular power spectrum of $T(\theta ,\varphi )$ than {\rm(\ref{SER})}.

\begin{theorem}\label{th4}
	Let $u(\theta ,\varphi ,t)$  be the solution to the initial value problem {\rm(\ref{TER1})-(\ref{TER3})} and  the angular power spectrum \{$C_{l},l=0,1,2,...\}$ of
	the random field $T(\theta ,\varphi )$ from the initial condition~{\rm(\ref{TER2})} satisfies the assumption 
	\[
	\sum_{l=0}^{\infty }(2l+1)^3C_{l}<\infty. 
	\]
	
	Then there exists a constant C such that for all $t>0$ it holds
	\[
	\|u(\theta ,\varphi ,t+h)-u(\theta ,\varphi ,t)\|_{L_2(\Omega \times
		{\mathbb S}^{2})} \le  Ch,\quad \mbox{when} \quad h\to 0+,
	\]
	where the constant $C$ depends only on the parameters $c$, $D$ and $k.$
\end{theorem}

Replacing $\sum_{l=0}^{\infty}$ by $\sum_{l=L}^{\infty}$ in the  proof of Theorem~\ref{th4} in Appendix B we obtain the H\"older continuity of the approximations to the solution.
\begin{corollary}
	Let the assumptions of Theorem~{\rm \ref{th4}} hold true. Then there exists a constant $C_L$ such that for all $t>0$ it holds
	\[
	\|u_L(\theta ,\varphi ,t+h)-u_L(\theta ,\varphi ,t)\|_{L_2(\Omega \times
		{\mathbb S}^{2})} \le  C_Lh,\quad \mbox{when} \quad h\to 0+,
	\]
	where the constant $C_L$ depends only on the parameters $c$, $D$ and $k.$
\end{corollary}

The following result is proven in Appendix B. It provides an upper bound on $p$th moments of the solution increments  in time.
\begin{corollary}\label{cor4}
	Let the assumptions of Theorem~{\rm \ref{th4}} hold true.	Then, for each $p>0,$ there exists a constant C such that for all $t>0$ it holds
	\[
	\|u(\theta ,\varphi ,t+h)-u(\theta ,\varphi ,t)\|_{L_p(\Omega \times
		{\mathbb S}^{2})} \le  Ch,\quad \mbox{when} \quad h\to 0+,
	\]
	where  the constant $C$ depends only on the parameters $p,$ $c$, $D$ and $k.$
\end{corollary}

Finally, we present continuity properties of the solution at time $t$ with respect to   the geodesic distance on the sphere.
\begin{corollary}\label{cor5}
	Let $u(\theta ,\varphi ,t)$  be the solution to the initial value problem {\rm(\ref{TER1})-(\ref{TER3})} and  the angular power spectrum \{$C_{l},l=0,1,2,...\}$ of
	the random field $T(\theta ,\varphi )$ from the initial condition~{\rm(\ref{TER2})} satisfies the assumption 
	\[
	\sum_{l=0}^{\infty }(2l+1)^{1+2\gamma}C_{l}<\infty, \quad\gamma \in[0,1].  \]
	
	Then, there exists a constant C such that for all $t>0$ it holds
	\[\mathbf{MSE}(u(\theta ,\varphi ,t)-u(\theta' ,\varphi' ,t)) \le  C
	\sum_{l=0}^{\infty}C_{l}\left(2l+1\right)^{1+2\gamma} (1-\cos\Theta)^\gamma,
	\]
	where $\Theta$ is the angular distance  between  $(\theta,\varphi)$ and $(\theta',\varphi')$ and the constant $C$ depends only on the parameters $c$, $D$ and $k.$
\end{corollary}

\section{Numerical studies}\label{sec6}
In this section, we present detailed numerical studies of the solution $u(\theta ,\varphi ,t)$  of the initial value problem {\rm(\ref{TER1})-(\ref{TER3})} and its approximation $u_L(\theta ,\varphi ,t).$  We investigate the convergence rates of the approximation to the solution and evolutions of the solution, the covariance function and its power spectrum over time.

We use the data with measurements of the CMB temperature intensity from Planck 2015 results, see \cite{PLANCK2} and \cite{PLANCK1}.
Figure~\ref{cmbmap} shows the CMB map produced from the SMICA (a component separation method for CMB data processing) pipeline data at $N_{\rm side}=1024$ at $10$ arcmin resolution with $12,582,912$ HEALPix points.  Figure~\ref{ang_spec} plots the best-fit LCDM scaled angular power spectrum $D_{l}=l(l+1)C_l/(2\pi),$ $l=2,..., 2508$, of the CMB map at the recombination time. The scaled CMB angular power $D_{l}$ is shown as a function of the harmonic number $l.$ It begins at $l=2$ as for $l=0$ and 1 it can not be reliably estimated using only $2l+1$ values.

We use the coefficients $a_{lm}$ and the angular power spectrum of CMB temperature intensities in Figures~\ref{cmbmap} and \ref{ang_spec}  as the initial condition of the Cauchy problem {\rm(\ref{TER1})-(\ref{TER3})}.  For numerical studies and R computations in this section we assume that the angular spectrum of the random field $T(\theta ,\varphi)$ is vanished if $l$ is greater than  2508. Thus, we use $u_{L_0}(\theta ,\varphi ,t)$ with $L_{0}=2508$ as a substitution of the solution $u(\theta ,\varphi ,t).$ Also, as the temperature of the ensemble of decoupled photons has continued to diminish and now shows very small variability in the range $2.7260\pm0.0013$~K, in plots we use the same colour scheme but different scales compared to the intensity map in Figure~\ref{cmbmap} that corresponds to time $t=0$ in the model {\rm(\ref{TER1})-(\ref{TER3})}.  It helps better visualise the solutions and their approximation errors.

{\bf \subsection{Evolution of solutions}}
Study of the evolution of CMB field is critical to unveil important properties of the present and primordial universe \cite{PLANCK2}, \cite{Dodelson}. The CMB map can be modelled as a realization of the random field  $u(\theta ,\varphi ,t)$ on $\mathbb{S}^2.$  We demonstrate its evolution due to the model  {\rm(\ref{TER1})-(\ref{TER3})}. For the following numerical examples we will use the random field $T(\theta ,\varphi)$  with the angular power spectrum of CMB temperature intensities given in Figure~\ref{ang_spec}.

\begin{figure}[th]\vspace{-5mm}
	\begin{minipage}{0.48\textwidth}
		\includegraphics[trim = 1mm 12mm 10mm 20mm, clip, width=1.0\textwidth,height=0.85\textwidth]{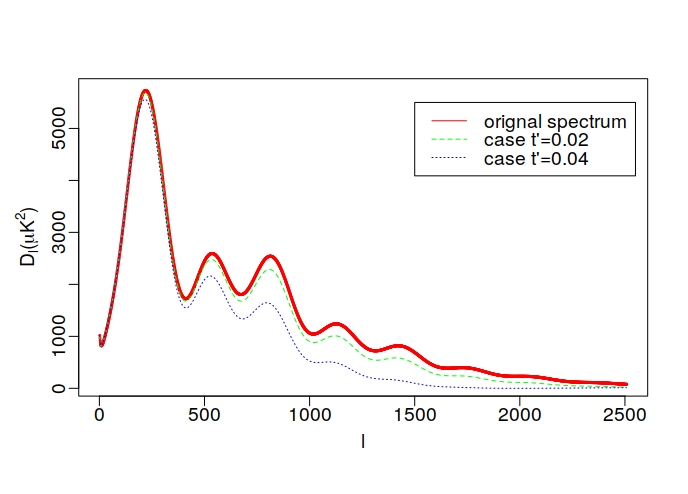}\\
		\caption{Scaled CMB angular power spectra for $c=1,$ $D=1$ and $k=0.01$ at time $t'=0,$ 0.02 and 0.04.}\label{angsp2}
	\end{minipage}\hspace{3mm}
	\begin{minipage}{0.48\textwidth}
		\includegraphics[trim = 1mm 5mm 10mm 13mm,clip, width=1\textwidth,height=0.97\textwidth]{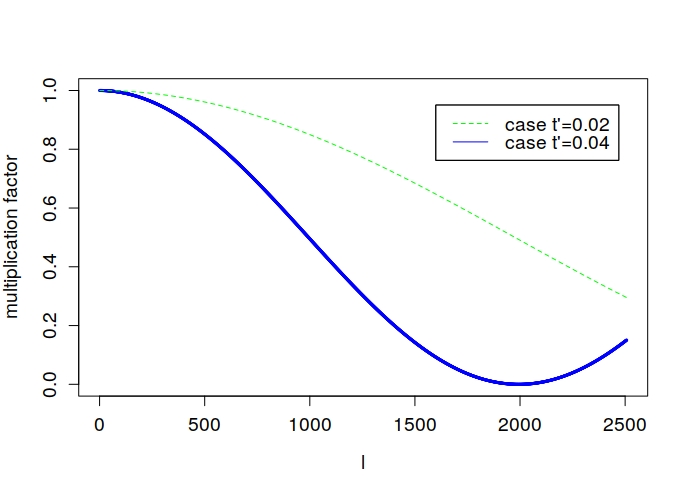}\\[-4mm]
		\caption{Multiplication factors for scaled CMB angular power spectra for $c=1,$ $D=1$ and $k=0.01$ at time $t'=0.02$ and 0.04.}\label{angsp3}
	\end{minipage}\vspace{-3mm}
\end{figure}

In the case of the model  {\rm(\ref{TER1})-(\ref{TER3})} it follows from (\ref{SOL}) and (\ref{xis}) that the angular spectrum evolution over time  is determined by the multiplication factor $\exp\left( -{c^{2}t}/{2D}\right) [A_{l}^2(t)+B_{l}^2(t)].$  Since the attenuation factor is $\exp(-c^2t/2D)$  the dimensionless time $t'=c^2t/2D$ will be used in all following plots.  Figure~\ref{angsp2} shows the original scaled angular power spectrum $D_{l}$ in red and  the angular power spectra at time $t'=0.02$ and 0.04 in green and blue respectively. It is observed that $D_{l}$ changes little from the original values over short periods. The deviations increase with increasing $l$ which is consistent with the changes of the multiplication factor shown in  Figure~\ref{angsp3} and cosmological theories showing that higher multipoles are changing faster.

\begin{figure}[bth]\vspace{-3mm}
	\begin{minipage}{0.45\textwidth}
		\includegraphics[trim = 2mm 5mm 10mm 10mm, clip, width=1\textwidth,height=1.13\textwidth]{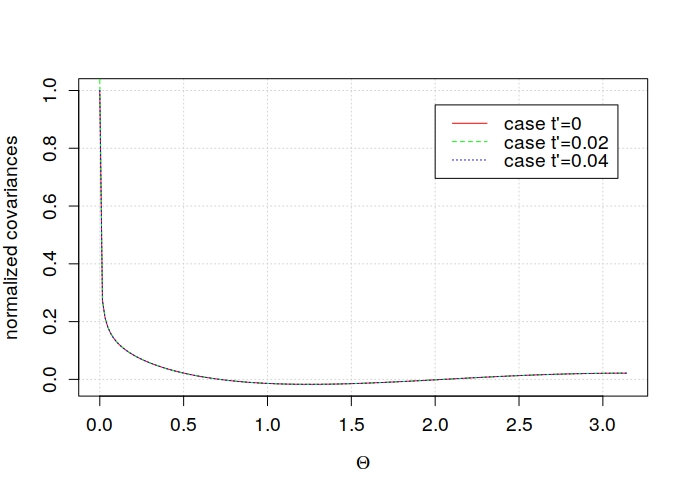}\\
		\caption{Three almost identical covariances at the time lags $t' =0,$ 0.02 and 0.04 at angular distances $\Theta$ for $c=1,$ $D=1$ and $k=0.01$.}\label{cov2d}
	\end{minipage}\hspace{3mm}
	\begin{minipage}{0.54\textwidth}
		\includegraphics[trim = 14mm 15mm 30mm 30mm,clip, width=1\textwidth,height=0.9\textwidth]{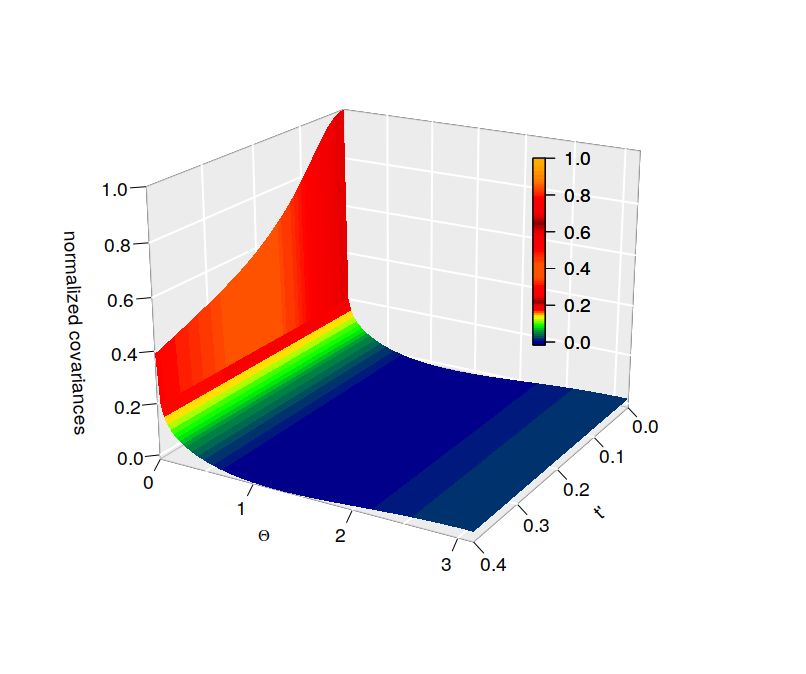}\\[-4mm]
		\caption{Covariance for $c=1,$ $D=1$ and $k=0.01$ at time lag $t'$ and angular distance $\Theta.$ }\label{cov3d}
	\end{minipage}\vspace{-4mm}
\end{figure}

Figure~\ref{cov2d} displays the covariance $\mathbf{Cov}(u(\theta ,\varphi ,0),u(\theta ^{\prime },\varphi ^{\prime },t^{\prime}))$ defined by (\ref{COV}). Three lines depict covariances at the time lags $t'=0,$ 0.02, and 0.04 as functions of  the
angular distance $\Theta.$ To further understand  the impact of time and the
angular distance on the covariance we produce 3d-plots showing the covariance as a function of the time lag $t'$ and the angular distance $\Theta,$ see Figure~\ref{cov3d}. 
The plots in Figures~\ref{cov2d} and \ref{cov3d} are normalised by dividing each value by the variance at time 0 and the angular distance 0, i.e. by $\mathbf{Cov}(u(0 ,0 ,0),u(0, 0,0)).$ It is observed that the covariance decays very rapidly. It changes very little over short time periods, except $\Theta$ values close to 0. As the angular power spectrum decreases very quickly only its values at small multipoles have the  principal impact on covariances. Relative large changes of $D_{l}$  at high frequencies (see Figure~\ref{angsp2}) do not substantially change the covariance function. It is evidenced from Figure~\ref{cov2d} where the three lines almost coincide. Hence, it would not be reasonable to use the covariance function to characterise fine changes in CMB maps over short periods of time. However, as the correlations change little over time it can help in studying the CMB at earlier epochs. 
\begin{figure}[h]\vspace{-7mm}
	\begin{minipage}{0.5\textwidth}
		\includegraphics[trim = 3mm 43mm 32mm 38mm, clip, width=1.05\textwidth]{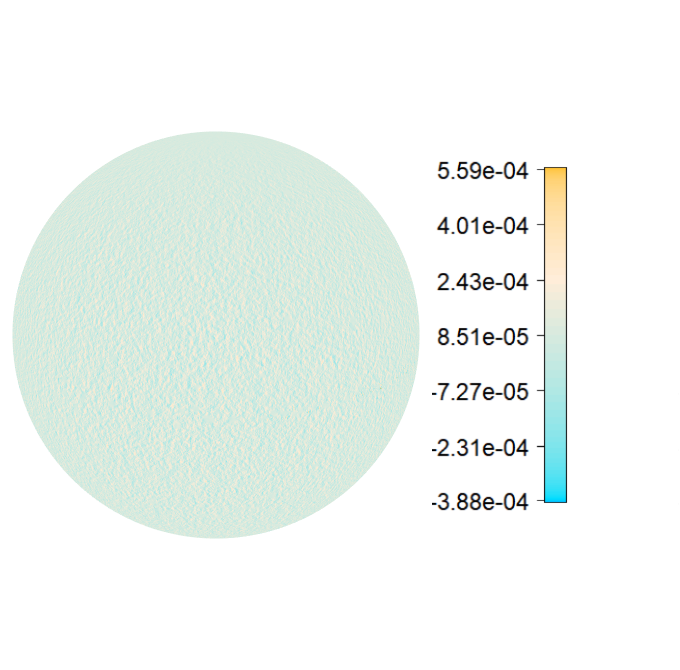}\\[3mm]
		\caption{Difference of  $u(\theta ,\varphi ,t')$ at time $t'=0$ and 0.04  for  $c=1,$ $D=1$ and $k=0.01.$}\label{cmbmap2dif}
	\end{minipage}\hspace{1mm}
	\begin{minipage}{0.49\textwidth}
		\includegraphics[trim = 1mm 10mm 32mm 18mm, clip, width=1\textwidth]{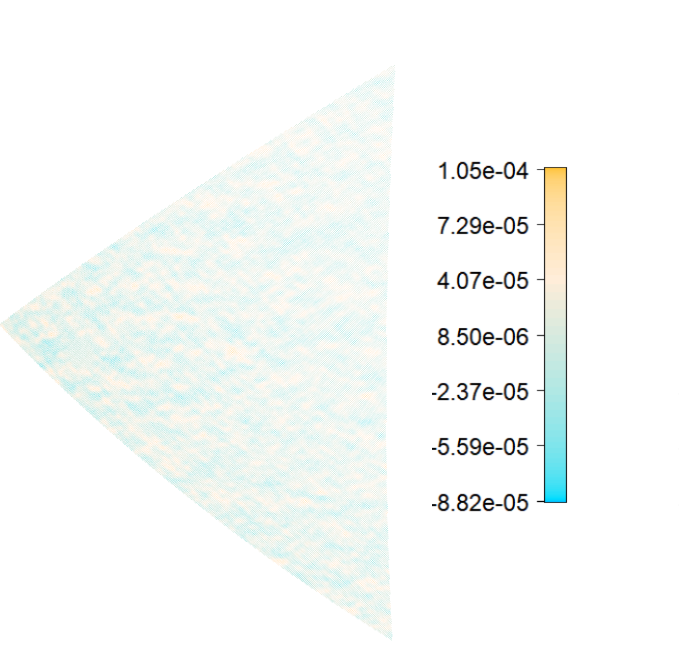}\\[-5mm]
		\caption{Difference of  $u(\theta ,\varphi ,t')$ at time \mbox{$t'=0$} and 0.04  for  $c=1,$ $D=1$ and $k=0.01$ in a small sky window.}\label{cmbmap2difsmall}
	\end{minipage}
\end{figure}

We investigated realisations of the solution that correspond to scaled angular power spectra for $c=1,$ $D=1$ and $k=0.01$ at early conformal times. Their temperature maps were rather similar to the original map in Figure~\ref{cmbmap} that corresponds to time $t'=0.$ The temperature field becomes smoother and its range narrows when time increases.   Figure~\ref{cmbmap2dif} depicts differences of the temperature intensities at time $t'=0$ and 0.04.  The fine structure of changes is shown in Figure~\ref{cmbmap2difsmall} that visualises a small triangular "equatorial" region around the Milky Way galaxy.   No regions with extreme changes compare to other locations can be seen in the both figures.

\begin{figure}[th]\vspace{-5mm}
	\begin{minipage}{0.5\textwidth}
		\includegraphics[trim = 3mm 43mm 32mm 38mm, clip, width=1\textwidth]{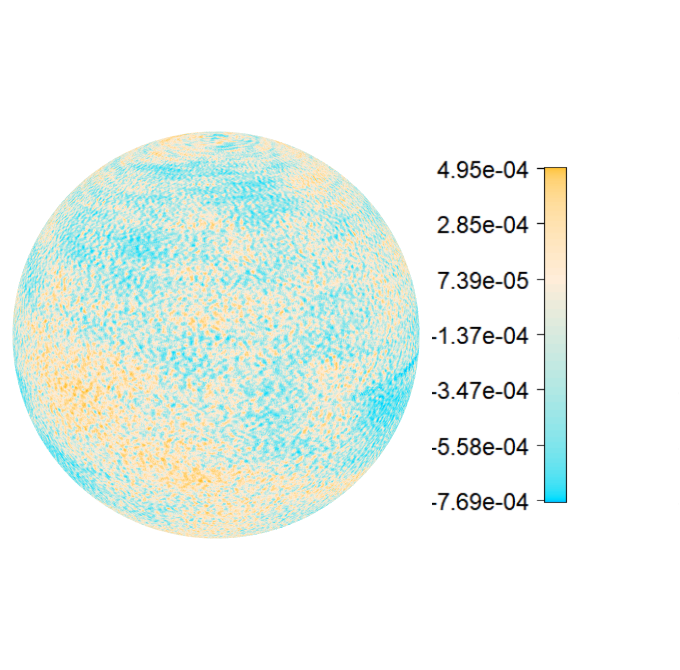}\\
		\caption{The realization of $u(\theta ,\varphi ,t')$ for $c=1,$ $D=1$ and $k=0.05$ at $t'=0.04.$}\label{cmbmap2k}
	\end{minipage}\hspace{1mm}
	\begin{minipage}{0.5\textwidth}
		\includegraphics[trim = 3mm 43mm 32mm 38mm, clip, width=1\textwidth]{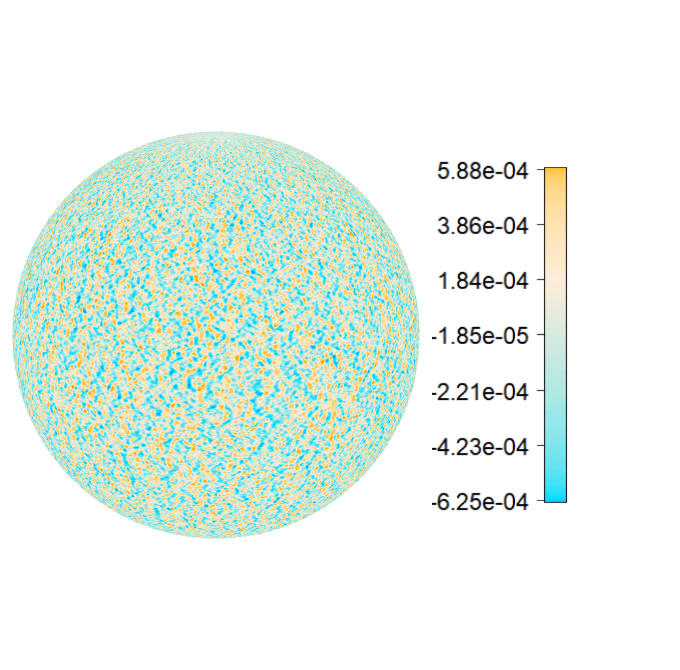}\\
		\caption{Difference of  $u(\theta ,\varphi ,t')$ for $k=0.01$ and $0.05$ when  $c=1,$ $D=1$ and $t'=0.04.$}\label{cmbmap2k1dif}
	\end{minipage}\vspace{-5mm}
\end{figure}

Figure~\ref{cmbmap2k} shows a realisation of the solution that corresponds to the case  $c=1,$ $D=1$ and $k=0.05$ at time $t'=0.04.$  The temperature intensity differences  between the solution fields with $k=0.01$ and $k=0.05$ (Figure~\ref{cmbmap2k}) at time $t'=0.04$ are visualised in Figure~\ref{cmbmap2k1dif}. As expected the higher value 0.05 of the diffusivity parameter $k$ results in a "blurred" realisation of the map with $k=0.01.$ The difference field does not exhibit any specific spatial pattern.

{\bf \subsection{Convergence rates of approximations to solutions}}
Now we analyse approximations to solutions and corresponding approximation errors depending on the truncation degree $L.$

We analysed the spatial error fields $u(\theta ,\varphi ,t')-u_{L}(\theta ,\varphi ,t')$ of approximations  $u_{200}(\theta ,\varphi ,0.04)$ and $u_{400}(\theta ,\varphi ,0.04)$  to the solution $u(\theta ,\varphi ,0.04)$ with $c=1,$ $D=1$ and $k=0.01$ at time $t'=0.04.$  The approximation error field for the case $L=200$ was rather similar to the true map which indicates that more terms are required to reconstruct fine details of the temperature intensity fields. For the case $L=400$ the error field  did not exhibit any specific spatial pattern. An increase in the approximation accuracy was also evidenced by the decrease of the mean squared error from $5.988295\cdot 10^{-09}$ to $4.075033\cdot 10^{-09}.$

\begin{figure}[h]\vspace{-5mm}
	\begin{center}
		\begin{minipage}{0.6\textwidth}
			\includegraphics[trim = 0mm 5mm 7mm 15mm,clip, width=1\textwidth,height=0.8\textwidth]{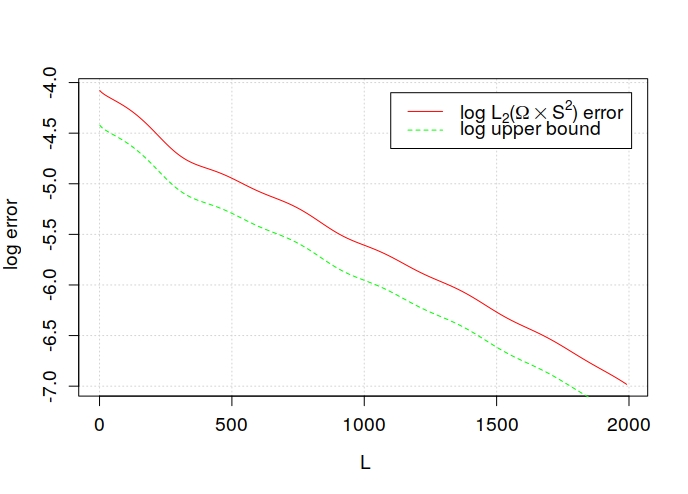}
			\caption{Logarithms of the mean ${L_2(\Omega \times
					{\mathbb S}^{2})}$-errors and their upper bound (\ref{upb0}) for $c=1,$ $D=1$ and $k=0.1$ at $t'=10.$}\label{errLt2}
		\end{minipage}
	\end{center}\vspace{-5mm}
\end{figure}

Figure~\ref{errLt2} shows the difference of the mean ${L_2(\Omega \times
{\mathbb S}^{2})}$ truncation errors $\Big\|u(\theta ,\varphi ,t')-u_L(\theta ,\varphi ,t')\Big\|_{L_2(\Omega \times
{\mathbb S}^{2})}$  and their upper bounds (\ref{upb0}) in Theorem~\ref{th3} on a natural logarithmic scale. The case of the SPDE (\ref{TER1}) with $c=1,$ $D=1$ and $k=0.1$ at $t'=10$ is considered. The difference is plotted as a function of~$L.$  
The plot confirms that both the error and its upper bound asymptotically vanish when $L$ increases. Moreover,  the convergence rates of the error and its upper bound are of the same order and differ only by a constant multiplication factor. Thus, the approximations to the solutions achieve an optimal order of convergence.  Figure~\ref{errLt2} also suggests that the convergence rate is faster than the power one.

Some additional numerical studies of the dependence on time and the parameters are given in Appendix C. The code in the folder "Research materials" from \mbox{\url{https://sites.google.com/site/olenkoandriy/}} can be used to obtain maps of realisations for different parameters that were not included in this section or to experiment with other combinations of the parameters.

\section{Conclusions and future work}\label{sec8}
This research provides physical motivation and justification for stochastic diffusion models in CMB studies. The Cauchy problem with random initial conditions for hyperbolic diffusion equations on the unit sphere was considered. Properties of exact and approximate solutions were investigated. The numerical studies illustrated the obtained results and examined the sensitivity of solutions to parameters using the CMB data. They indicated that the model is flexible enough to capture some of the statistical properties of the CMB.

It was shown that properties of the solutions are determined by the decay of the angular power spectrum. The analysis of the CMB covariance revealed that dependencies between CMB observations rapidly decrease with angular distance between their locations. The numerical studies demonstrated that the CMB  evolution under this SPDE model results in most significant changes of CMB  temperature spectrum at the high frequency range in the first place.  At the same time relatively large changes of $D_{l}$ at high CMB frequencies have little impact on the covariance functions and CMB maps that remain almost unchanged. The contribution of high frequency components to the CMB field  is decreasing with $l$ at a faster rate than a power-law.

The sensitivity analysis to parameters  at the lower frequency spectrum range demonstrated that changes in time and the parameters  $c$ and $D$ have a substantial impact on the error, while $k$ showed almost no effect. 

It would be interesting to extend the obtained results and  
\begin{itemize}
	\item[(1)]  apply this model to backward studies of CMB;
	\item[(2)] compare this model with other possible evolution scenarios;
	\item[(3)] investigate 3-dimensional SPDEs as the next generation CMB experiment, CMB-S4, will be collecting 3D spatial data;
	\item[(4)] explore applications to other spherical data in physics and earth sciences.
\end{itemize}

\begin{acknowledgements} This research was supported under the Australian
Research Council's Discovery Project DP160101366. N. Leonenko was supported in part by Cardiff
Incoming Visiting Fellowship Scheme, International Collaboration Seedcorn
Fund, Data Innovation URI Seedcorn Fund. We are also grateful for the use of data of the
Planck/ESA mission from the Planck Legacy Archive. The authors are also grateful to the referees for their careful reading of the paper and suggestions that helped to improve the paper.
\end{acknowledgements}

\section*{Appendix A: Diffusion length of a local disturbance.}
Consider a density disturbance $u$ of total mass $Q$ originating at the origin.  The well-known point source solution to linear diffusion in three space dimensions is given by
\[
u=\frac{1}{8[\pi Dt]^{3/2}}e^{-r^2/4Dt}.
\]
The density level set at some low significance value $u$ is at
\[
r=2[Dt]^{1/2}\ln ^{1/2}\left(\frac{Q}{u[8\pi Dt]^{3/2}}\right),\]
so 
\[\frac{dr}{dt}=(D/t)^{1/2}\frac{\ln \left(\frac{Q}{u[8\pi Dt]^{3/2}}\right)-3/2}{\ln ^{1/2}\left(\frac{Q}{u[8\pi Dt]^{3/2}}\right)}.
\]
The level set reaches its maximum extent when $\frac{dr}{dt}=0$, implying 
$t=\frac{(Q/u)^{2/3}}{8e\pi D},$
so the diffusion length is 
\[ r_D=\frac 12\left(\frac{3}{\pi e}\right)^{1/2} \left(\frac Qu\right)^{1/3}\approx 0.296  \left(\frac Qu\right)^{1/3}.
\]
For example, for a mass disturbance the size of a solar mass, and a neutron diffusion length of 0.3 light-year at the temperature of neutrino dissociation from weak nuclear interactions, estimated from \cite{Applegate} and \cite{Kurki}, the marginal disturbance density $u$ is around 1 solar mass per cubic light year. This is meant to have occurred at a time when the cosmological expansion factor $a(t)$ was less than $10^{-3}$, so after expansion to the current level, the equivalent marginal density would be less than one nucleon mass per cubic metre, around the current mean density of the universe.

\section*{Appendix B: Proofs.}

\noindent{\bf Proof of Theorem~{\rm\ref{Th1}}.}
By substituting (\ref{series}) into equation (\ref{teleq}) and using (\ref{sphharm}), we
obtain 
\begin{equation}
\sum_{l=0}^{\infty }\sum_{m=-l}^{l}\left[ \frac{1}{c^{2}}\frac{{ d} 
	^{2}b_{lm}(t)}{{ d}\, t^{2}}+\frac{1}{D}\frac{{d}\, b_{lm}(t)}{{ d}\,
	t}+l(l+1)k^{2}b_{lm}(t)\right] Y_{lm}(\mathbf{x})=0.  \label{series1}
\end{equation}%
To find particular solutions of (\ref{series1}), we need to solve the
ordinary differential equation 
\begin{equation}
\frac{1}{c^{2}}\frac{{ d}^{2}b_{lm}(t)}{{ d}\, t^{2}}+\frac{1}{D}%
\frac{{d}\, b_{lm}(t)}{{d}\, t}+l(l+1)k^{2}b_{lm}(t)=0.  \label{ODEq}
\end{equation}%
The initial conditions for this equation can be determined from (\ref{coeff}%
) and (\ref{incond}) and they are 
\begin{equation}
b_{lm}(t)|_{t=0}= \tilde{Y}_{lm}^{\ast }(\mathbf{0}),\qquad \left. \frac{%
	{ d}\, b_{lm}(t)}{{ d}\, t}\right\vert _{t=0}=0.  \label{condA}
\end{equation}%
The characteristic equation of (\ref{ODEq}) is
$
\frac{1}{c^{2}}z^{2}+\frac{1}{D}z+l(l+1)k^{2}=0,  \label{CharEq}\nonumber
$
with the roots $z_{1,2}=-{c^{2}}/(2D)\pm K_l. $  
Therefore, the general solution of equation (\ref{ODEq}) is given by the formula: 
\[
b_{lm}(t)=M_{1}e^{z_{1}t}+M_{2}e^{z_{2}t}, 
\]%
where $M_{1},M_{2}$ are some constants. From the initial conditions in (\ref{condA}%
) we obtain 
\[
M_{1}  =  \left( \frac{1}{2}+\frac{c^{2}}{4DK_l}\right) \tilde{Y}_{lm}^{\ast }(\mathbf{0}) , \quad M_{2} =  \left( \frac{1}{2}-\frac{c^{2}}{4DK_l}\right) \tilde{Y}_{lm}^{\ast }(\mathbf{0}). 
\]

Thus, the solution of the Cauchy problem (\ref{ODEq})-(\ref{condA}) is given by
\[
 b_{lm}(t)  =  \left( \frac{1}{2}+\frac{c^{2}}{4DK_l}\right) \tilde{Y}_{lm}^{\ast }(\mathbf{0}) \exp \left[ -t\left( \frac{c^{2}}{2D}-K_l\right) \right] \]
\[ +  \left( \frac{1}{2}-\frac{c^{2}}{4DK_l}\right) \tilde{Y}_{lm}^{\ast }(\mathbf{0})  \exp \left[ -t\left( \frac{c^{2}}{2D}+K_l\right) \right] . \]

Returning now to (\ref{series}), we obtain the solution of the Cauchy problem (%
\ref{teleq})-(\ref{incond}) in the form
\[ \hspace{-10mm}\tilde{p}(\mathbf{x},t) =    \sum_{l=0}^{\infty }Q_{l}(\mathbf{x})  \left(\left( \frac{1}{2}+\frac{c^{2}}{4DK_l}\right) \exp \left[ -t\left( \frac{c^{2}}{2D} - K_l\right) \right] \right.\]
\begin{equation} +   \left. \left( \frac{1}{2}-\frac{c^{2}}{4DK_l}\right) \exp \left[ -t\left( \frac{c^{2}}{2D} + K_l\right) \right]\right). \label{FinalSolution}
\end{equation}

Note that the multiplier of $Q_{l}(\mathbf{x})$ on the right-hand side of (\ref{FinalSolution}) equals 
\[ \exp\left( -\frac{c^{2}t}{2D}\right) \biggl\{ \cosh \left( tK_l\right)  + \frac{c^{2}}{2DK_l}%
\;\sinh \left( tK_l\right) %
\biggr\}. \]

By substituting this expression into (\ref{FinalSolution}), we get 
\[
\tilde{p}(\mathbf{x},t) = \exp\left(-\frac{c^{2}t}{2D}%
\right) \sum_{l=0}^{\infty} Q_{l}(\mathbf{x})\biggl\{ \cosh \left( tK_l\right) 
+ \frac{c^{2}}{2DK_l}%
\;\sinh \left( t K_l\right) %
\biggr\} . \]

Finally, using $K_l'$ and rewriting the Green function we obtain the statement of the theorem.

\

\noindent{\bf Proof of Theorem~{\rm\ref{Th2}}.}
The solution of the initial value problem (\ref{TER1}) - (\ref{TER3}) can
be written as a spherical convolution of the Green function $p(\theta ,\varphi
,t)$ from Section~\ref{sec4} and the random field $T(\theta ,\varphi ),$ if the corresponding Laplace
series converges in the Hilbert space $L_{2}(\Omega \times \mathbb{S}^2,\sin \theta
d\theta d\varphi ).$

Let the two functions $f_1(\cdot)$ and $f_2(\cdot)$
on the sphere $\mathbb{S}^2$ belong to the space $L_{2}(\mathbb{S}^2,\sin\theta d\theta
d\varphi )$ and have the Fourier-Laplace coefficients 
\[
a_{lm}^{(i)}=\int_{\mathbb{S}^2}f_{i}(\theta ,\varphi )Y_{lm}^{\ast }(\theta
,\varphi )\sin \theta d\theta d\varphi , \qquad i=1,2. 
\]%
Recall (see, i.e., \cite{Dunkel}) that  their non-commutative spherical convolution  is defined as the Laplace series 
\begin{equation}
\lbrack \ f_{1}\ast \ f_{2}](\theta ,\varphi )=\sum_{l=0}^{\infty
}\sum_{m=-l}^{l}a_{lm}^{(\ast )}\ Y_{lm}(\theta ,\varphi ) \label{SERC}
\end{equation}%
with the Fourier-Laplace coefficients given by
\[
a_{lm}^{(\ast )}= \sqrt{\frac{4\pi }{2l+1}}a_{lm}^{(1)}a_{l0}^{(2)}, 
\]%
provided that the series (\ref{SERC}) converges in the corresponding Hilbert
space.

Thus, the random solution $u(\theta ,\varphi ,t)$ of equation (\ref{TER1}) with the initial values determined by (\ref{TER2}) and (\ref{TER3}) can be written as a spherical random field
with the following Laplace series representation 
\begin{equation}\label{SER1}
u(\theta ,\varphi ,t)=[\ T\ast \ p_{t}](\theta ,\varphi )=\sum_{l=0}^{\infty
}\sum_{m=-l}^{l}a_{lm}^{(t)}Y_{lm}(\theta ,\varphi ),  
\end{equation}%
provided that this series is convergent in the Hilbert space $L_{2}(\Omega \times
\mathbb{S}^2,\sin \theta d\theta d\varphi ),$ where $p_{t}=p(\theta ,\varphi ,t)$
is given by Theorem~\ref{Th1} and $T$ is given by (\ref{TER2}). 
 The complex Gaussian random variables $a_{lm}^{(t)}$ are given by 
\[
a_{lm}^{(t)}= \sqrt{\frac{4\pi }{2l+1}}a_{lm}a_{l0}^{(p_{t})}, 
\]%
where $a_{l0}^{(p_{t})}=Y_{l0}^{\ast }(\mathbf{0})d_{l}(\theta ,\varphi ,t)$ and 
\[
 d_{l}(\theta ,\varphi ,t)  =   \exp\left( -\frac{c^{2}t}{2D}\right) %
	\biggl\{\biggl[ \cosh \left( tK_l%
	\right) + \frac{c^{2}}{2DK_l}\;\sinh \left( t%
	K_l\right) \biggr] \]
\[\times \mathbf{1}%
	_{\left\{ l\leq \frac{\sqrt{D^{2}k^{2}+c^{2}}-Dk}{2Dk}\right\} }
 +  \biggl[ \cos \left( tK_l'\right) +\frac{c^{2}}{2DK_l'} \sin \left( t%
	K_l'\right) \biggr] \mathbf{1}%
	_{\left\{ l>\frac{\sqrt{D^{2}k^{2}+c^{2}}-Dk}{2Dk}\right\} } \biggr\}. 
\]
It gives the first statement of the theorem. 

By the addition formula for spherical harmonics (see, i.e., \cite%
{MarinucciPeccati13}, p.66)
\begin{equation}\label{asSp}
\sum_{m=-l}^{l}Y_{lm}(\theta ,\varphi )Y_{lm}^{\ast }(\theta ^{\prime
},\varphi ^{\prime }) = \frac{2l+1}{4\pi} P_l(\cos\Theta), 
\end{equation}%
where $P_l(\cdot)$ is the $l$-th Legendre polynomial (see (\ref{2.3})), and $\cos\Theta$ is
the angular distance between the points $(\theta,\varphi)$ and $%
(\theta^{\prime},\varphi^{\prime})$ on $\mathbb{S}^2.$

Using (\ref{Yl0}) we obtain that  the random field $u(\theta ,\varphi ,t)$
is isotropic if and only if the covariance structure of the solution (\ref%
{SOL}) can be written in the form 

\[ \mathbf{Cov}(u(\theta ,\varphi ,t),u(\theta ^{\prime },\varphi ^{\prime },t^{\prime}))  = \exp\left( -\frac{c^{2}}{2D}(t+t^{\prime })\right)\]
\[\times 
\sum_{l=0}^{\infty}\sum_{m=-l}^{l}Y_{lm}(\theta ,\varphi )Y_{lm}^{\ast
}(\theta ^{\prime },\varphi ^{\prime }) \mathbf{E}\xi _{lm}(t)\xi^{\ast
} _{lm}(t^{\prime }),\]
which gives the result in (\ref{COV})
provided the series (\ref{COV}) converges for every fixed $t$ and $t^{\prime
},$ that is 
\begin{equation}
\sum_{l=0}^{\infty }(2l+1)C_{l}
P_l(\cos\Theta) [A_{l}(t)A_{l}(t^{\prime })+B_{l}(t)B_{l}(t^{\prime
})]<\infty .  \label{COV1}
\end{equation}%

Noting that  $\left\vert P_l(\cos\Theta)\right\vert\le 1,$ only a finite number of terms $A_{l}$ is non-zero,   and there is a constant $C$ such that $\sup_{t\ge 0} |B(t)|<C,$ we obtain that condition (\ref{COV1}) follows from~(\ref{SER}). 
This condition on the angular spectrum $C_{l}, l\geq 0,$ guarantees the convergence of the series~(\ref{SER1})  in the Hilbert space $L_{2}(\Omega \times
{\mathbb S}^{2},\sin\theta d\theta d\varphi ).$

\

\noindent{\bf Proof of Theorem~{\rm\ref{th3}}.}
	The approximation $u_L(\theta ,\varphi ,t)$	is a centered Gaussian random field, i.e. $ \mathbf{E}u_L(\theta ,\varphi ,t)=0$ for all $L\in\mathbb{N},$ $\theta \in \lbrack 0,\pi ),$ $\varphi \in \lbrack 0,2\pi ),$ and $t>0.$ Therefore,
	\[
	\Big\|u(\theta ,\varphi ,t)-u_L(\theta ,\varphi ,t)\Big\|_{L_2(\Omega \times
		{\mathbb S}^{2})} 
	= \exp\left( -\frac{c^{2}t}{2D}\right)\]
		\[\times
	\left(\sum_{l=L}^{\infty}\sum_{m=-l}^{l}Y_{lm}(\theta ,\varphi )Y_{lm}^{\ast
	}(\theta ,\varphi ) \mathbf{E}\xi _{lm}(t)\xi^{\ast
	}_{lm}(t)\right)^{1/2} \]
		\begin{equation}=\frac{1}{2\sqrt{\pi}}\exp\left( -\frac{c^{2}t}{2D}\right)
	\left(\sum_{l=L}^{\infty}(2l+1)C_{l}\cdot
	[A_{l}^2(t)+B_{l}^2(t)]\right)^{1/2}.\label{upb}
	\end{equation}
	
	By (\ref{At}) and (\ref{Bt}) we get
	\begin{equation}|A_{l}(t)|\le C\exp\left( \frac{c^{2}t}{2D}\right)\quad \mbox{and} \quad \sup_{t\ge 0}|B_{l}(t)|\le C.\label{upbBt}
	\end{equation}
	Hence, for all $L\in\mathbb{N}$ it holds
	
	\[\|u(\theta ,\varphi ,t)-u_L(\theta ,\varphi ,t)\|_{L_2(\Omega \times
		{\mathbb S}^{2})}\le C
	\left(\sum_{l=L}^{\infty}(2l+1)C_{l}\right)^{1/2}.\]
	
	For $l>\frac{\sqrt{D^{2}k^{2}+c^{2}}-Dk}{2Dk}$ it follows from (\ref{At}) that $A_{l}(t)\equiv 0$. Therefore, by (\ref{upb}) and (\ref{upbBt}) we obtain
	\[\|u(\theta ,\varphi ,t)-u_L(\theta ,\varphi ,t)\|_{L_2(\Omega \times
		{\mathbb S}^{2})}\le C\exp\left( -\frac{c^{2}t}{2D}\right)
	\left(\sum_{l=L}^{\infty}(2l+1)C_{l}\right)^{1/2}.\]

\

\noindent{\bf Proof of Corollary~{\rm\ref{cor2}}.}
	The statement (i) immediately follows from (\ref{upb0}) and the estimate
	\[\sum_{l=L}^{\infty}(2l+1)C_{l}\le C\sum_{l=L}^{\infty}l^{-(\alpha-1)}=C
	{L}^{-(\alpha-2)}.\]
	
	Then, applying Chebyshev's inequality, we get the upper bound in (ii).
	
	Finally, (iii) follows from statement (ii) and the Borel–Cantelli lemma as
	\[\sum_{l=L}^{\infty} \frac{1}{L^{\alpha-2}L^{-2\beta}}<\infty.\]
	
\

\noindent{\bf Proof of Theorem~{\rm\ref{th4}}.} Let $h$ belong to a bounded neighbourhood of the origin. It follows from (\ref{alm}), (\ref{xis}), (\ref{At}), (\ref{Bt}) and  (\ref{asSp}) that
	\[
	\Big\|u(\theta ,\varphi ,t+h)-u(\theta ,\varphi ,t)\Big\|_{L_2(\Omega \times
		{\mathbb S}^{2})} = \left\| \exp\left(-\frac{c^{2}(t+h)}{2D}\right)
	\sum_{l=0}^{\infty}\sum_{m=-l}^{l}Y_{lm}(\theta ,\varphi )\right.\]
	\[\left.\times\,\xi _{lm}(t+h)- \exp\left(-\frac{c^{2}t}{2D}\right)
	\sum_{l=0}^{\infty}\sum_{m=-l}^{l}Y_{lm}(\theta ,\varphi )\xi _{lm}(t)\right\|_{L_2(\Omega \times
		{\mathbb S}^{2})}\]
	\[=\frac{1}{2\sqrt{\pi}}\exp\left( -\frac{c^{2}t}{2D}\right)
	\left(\sum_{l=0}^{\infty}(2l+1)C_{l}\right.
	\left[\left(\exp\left( -\frac{c^{2}h}{2D}\right)A_{l}(t+h)-A_{l}(t)\right)^2\right.\]
		\begin{equation}\label{ABsum}\left.\left.+\left(\exp\left( -\frac{c^{2}h}{2D}\right)B_{l}(t+h)-B_{l}(t)\right)^2\right]\right)^{1/2}.
	\end{equation}
	
	We start by showing how to estimate the first summand in (\ref{ABsum}). By (\ref{At}), for the case $l=0$ we obtain
	\[\left(\exp\left( -\frac{c^{2}h}{2D}\right)A_{0}(t+h)-A_{0}(t)\right)^2\]
	\[=\left(\exp\left( -\frac{c^{2}h}{2D}\right) \exp\left(\frac{c^{2}(t+h)}{2D}\right)\right.\left.-\exp\left(\frac{c^{2}t}{2D}\right)\right)^2=0.\]
	For $l>0$ we will use the upper bound
	\[\left(\exp\left( -\frac{c^{2}h}{2D}\right)A_{l}(t+h)-A_{l}(t)\right)^2=\left(\exp\left( -\frac{c^{2}h}{2D}\right)\left(A_{l}(t+h)-A_{l}(t)\right)\right.\]
	\[\left.-\left(1-\exp\left( -\frac{c^{2}h}{2D}\right)\right)A_{l}(t)\right)^2\le 2\left(A_{l}(t+h)-A_{l}(t)\right)^2\]
	\[+2\left(1-\exp\left( -\frac{c^{2}h}{2D}\right)\right)^2A^2_{l}(t).\]
	By properties of $\cosh(\cdot)$ and $\sinh(\cdot)$  we get
	\begin{eqnarray}\cosh(x)-\cosh(y)&=&\frac{\exp{(x)}}{2}
	\left(1-\exp\left(-(x+y)\right)\right)\left(1-\exp\left(-(x-y)\right)\right),\nonumber\\
	\sinh(x)-\sinh(y)&=&\frac{\exp{(x)}}{2}\left(1+\exp\left(-(x+y)\right)\right)\left(1-\exp\left(-(x-y)\right)\right).\nonumber
	\end{eqnarray}
	
	Then, applying (\ref{At})  and noting that only a finite number of $A_l$ is non-vanished (namely, only if $l\in \left[0, \frac{\sqrt{D^{2}k^{2}+c^{2}}-Dk}{2Dk}\right]$) we obtain the following estimates 
	\[\left(A_{l}(t+h)-A_{l}(t)\right)^2\le \frac{\exp \left( 2(t+h)K_l\right)}{2}\Big[\left(1- \exp\left( -(t+h/2)K_l\right)\right)^2\]
		\[\times\left(1- \exp\left( -h K_l/2\right)\right)^2+\frac{c^4}{4D^2K_l^2}\left(1+ \exp\left( -(t+h/2)K_l\right)\right)^2\]
					\[\times\left(1- \exp\left( -h K_l/2\right)\right)^2\Big]\le C\exp \left( 2hK_l\right)\exp \left( 2tK_l\right) \]
					\[\times\left(1- \exp\left( -h K_l/2\right)\right)^2\le C\exp \left( 2tK_l\right)h^2,\]
	\[\left(1-\exp\left( -\frac{c^{2}h}{2D}\right)\right)^2A^2_{l}(t)\le\frac{c^{4}}{4D^2}h^2A^2_{l}(t)=\frac{c^{4}}{8D^2}h^2\exp \left( 2tK_l\right)\]
	\[\times\Big[\left(1+ \exp\left( -2tK_l\right)\right)^2+\frac{c^4}{4D^2K_l^2}\left(1- \exp\left( -2tK_l\right)\right)^2\Big]\le C\exp \left( 2tK_l\right)h^2.\]

	Now we estimate the second summand in (\ref{ABsum}) as
	\[\left(\exp\left(-\frac{c^{2}h}{2D}\right)B_{l}(t+h)-B_{l}(t)\right)^2\le \left(\exp\left( -\frac{c^{2}h}{2D}\right)\left(B_{l}(t+h)-B_{l}(t)\right)\right.\]
	\[\left.-\left(1-\exp\left( -\frac{c^{2}h}{2D}\right)\right)B_{l}(t)\right)^2\le 2\left(B_{l}(t+h)-B_{l}(t)\right)^2\]
	\[+2\left(1-\exp\left( -\frac{c^{2}h}{2D}\right)\right)^2B^2_{l}(t).\]
	
	Using (\ref{upbBt}) and applying the inequalities $|\cos(x)-\cos(y)|\le 2\left|\sin\left(\frac{x-y}{2}\right)\right|\le |x-y|$ and $|\sin(x)-\sin(y)|\le |x-y|$ we obtain
	\[\left(B_{l}(t+h)-B_{l}(t)\right)^2\le 2 \left((K_l')^2+\frac{c^4}{4D^2}\right) h^2,\]
	\[\left(1-\exp\left( -\frac{c^{2}h}{2D}\right)\right)^2B^2_{l}(t)\le
	\left(1-\exp\left( -\frac{c^{2}h}{2D}\right)\right)^2\left(1+\frac{c^2}{2D}\right)^2\le C h^2.\]
	Note that for all $l\ge 0$ it holds
	\[K_l\le \frac{c^{2}}{2D}\quad \mbox{and}\quad K_l'\le C(2l+1).\] 
	
	Applying the above estimates to (\ref{ABsum}) we obtain
	
	\[
	\Big\|u(\theta ,\varphi ,t+h)-u(\theta ,\varphi ,t)\Big\|_{L_2(\Omega \times{\mathbb S}^{2})} \le C\exp\left( -\frac{c^{2}t}{2D}\right)
	\left(\sum_{l=0}^{\infty}(2l+1)C_{l}\right.\]
		\[\left.\times\Big[\exp \left( 2tK_l\right)
	+(K_l')^2+C\Big]\right)^{1/2}h\le C\left(\sum_{l=0}^{\infty}(2l+1)^3C_{l}\right)^{1/2}h,\]
	which completes the proof.

\

\noindent{\bf Proof of Corollary~{\rm\ref{cor4}}.}
	Note that $u(\theta ,\varphi ,t)$	is a centered Gaussian random field and for any centered Gaussian random variable $X$ it holds 
	\[\mathbf{E}|X|^{p}= \frac{2^{p/2}\Gamma \left({\frac {p+1}{2}}\right)}{\sqrt{\pi }}\left(\mathbf{E}|X|^{2}\right)^{p/2}.\]
	Applying this result to the statement of Theorem~\ref{th4} we obtain
	\[\|u(\theta ,\varphi ,t+h)-u(\theta ,\varphi ,t)\|_{L_p(\Omega \times
		{\mathbb S}^{2})} = C \|u(\theta ,\varphi ,t+h)-u(\theta ,\varphi ,t)\|_{L_2(\Omega \times
		{\mathbb S}^{2})}\le  Ch.\]	

\		

\noindent{\bf Proof of Corollary~{\rm\ref{cor5}}.}
 By(\ref{COV})  it holds
	\[
	\mathbf{Var}\left(u(\theta ,\varphi ,t)-u(\theta',\varphi',t)\right)=\mathbf{Var}\left(u(\theta ,\varphi ,t)\right)\]
	\[+\mathbf{Var}\left(u(\theta',\varphi',t)\right)-2\,\mathbf{Cov}(u(\theta ,\varphi ,t),u(\theta ',\varphi ',t))\]
	\[= C\exp\left( -\frac{c^{2}t}{D}\right)
	\sum_{l=0}^{\infty}C_{l}\left(2l+1\right)
	\left(A_{l}^2(t)+B^2_{l}(t)\right)(1-P_l(\cos\Theta)).\]
	
	Applying the next property of Legendre polynomials (see, for example, \cite{Lang}, p.16) $|1 - P_l (x)| \le 2|1 - x|^\gamma (l(l + 1))^\gamma, \quad \gamma\in [0,1],$
	and the upper bounds (\ref{upbBt}), we obtain that uniformly in $t\ge 0$
	\[
	\mathbf{Var}\left(u(\theta ,\varphi ,t)-u(\theta',\varphi',t)\right)\le 
	C
	\sum_{l=0}^{\infty}C_{l}\left(2l+1\right)^{1+2\gamma} (1-\cos\Theta)^\gamma.
	\]	
\section*{Appendix C: Sensitivity to parameters.}
\begin{figure}[h]\vspace{-2mm}
		\begin{minipage}{0.5\textwidth}
			\includegraphics[trim = 20mm 14mm 35mm 20mm, clip,
			 width=1.0\textwidth,height=0.9\textwidth]{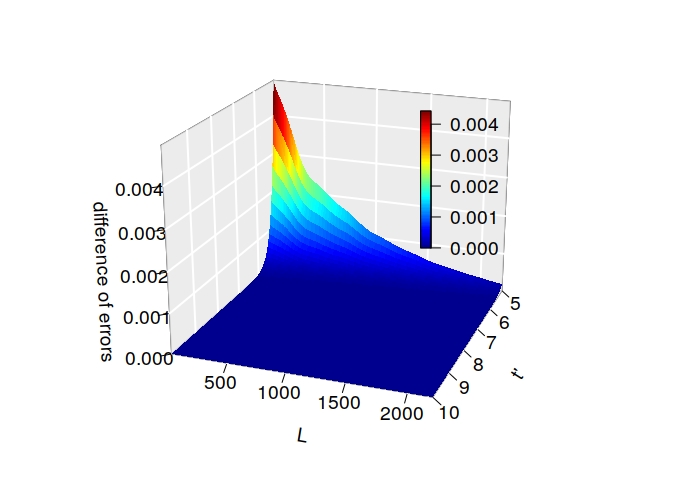}\\
			\caption{Difference of the mean ${L_2(\Omega \times
					{\mathbb S}^{2})}$-errors and their upper bound (\ref{upb0}) for $c=1,$ $D=1$ and $k=0.1.$}\label{errLt3}
		\end{minipage}\hspace{2mm}
		\begin{minipage}{0.5\textwidth}
			\includegraphics[trim = 20mm 14mm 35mm 20mm,clip, width=1\textwidth,height=0.9\textwidth]{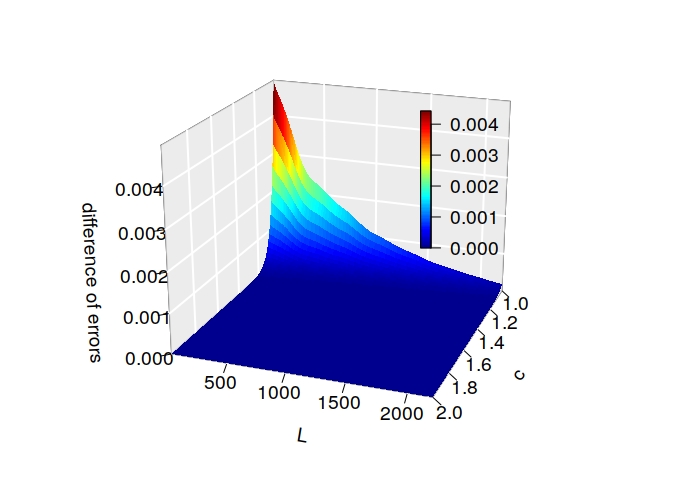}\\
			\caption{Difference of the mean ${L_2(\Omega \times
							{\mathbb S}^{2})}$-errors and their upper bound (\ref{upb0}) for $D=1$ and $k=0.1$ at $t'=10.$}\label{errLc3}
		\end{minipage}\\[5mm]
			\begin{minipage}{0.49\textwidth}
				\includegraphics[trim =  20mm 14mm 35mm 20mm, clip, width=1.0\textwidth,height=0.9\textwidth]{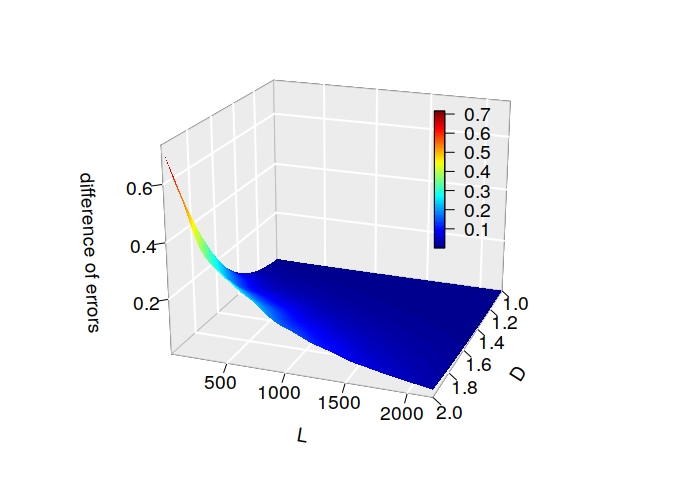}\\
				\caption{Difference of the mean ${L_2(\Omega \times
						{\mathbb S}^{2})}$-errors and their upper bound (\ref{upb0}) for $c=1$ and $k=0.1$ at $t'=10.$}\label{errLD3}
			\end{minipage}\hspace{1mm}
			\begin{minipage}{0.49\textwidth}
				\includegraphics[trim =  20mm 14mm 35mm 20mm,clip, width=1\textwidth,height=0.9\textwidth]{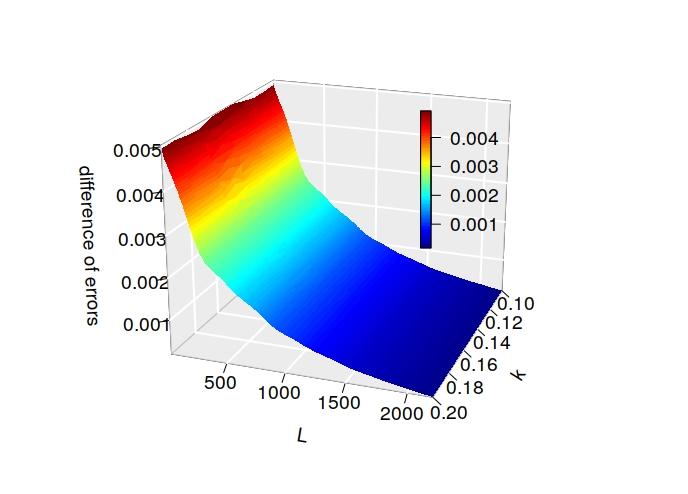}\\
				\caption{Difference of the mean ${L_2(\Omega \times
								{\mathbb S}^{2})}$-errors and their upper bound (\ref{upb0}) for $c=1$ and $D=1$ at $t'=10.$}\label{errLk3}
			\end{minipage}
		\end{figure}

	To further understand  the impact of time and the model parameters on the difference of the mean ${L_2(\Omega \times {\mathbb S}^{2})}$-errors and their upper bound (\ref{upb0}) we produced 3d-plots showing the difference as a function of the truncation degree $L$ and each parameter provided that other parameters are fixed.  These plots are displayed in Figures~\ref{errLt3}, \ref{errLc3}, \ref{errLD3}, and \ref{errLk3}. 

	In all cases the difference between the error and its upper bound asymptotically vanish when $L$ increases. Figure~\ref{errLt3} demonstrates that the difference is a decreasing function of time $t',$ which is expected as the series representation (\ref{SOL}) of the solutions $u(\theta ,\varphi ,t')$ has the multiplication factor $\exp\left(-t'\right)=\exp\left(-{c^{2}t}/(2D)\right)$  exponentially decaying in time. 
	 The differences are extreme at the origin and decrease when time or the parameter $c$ increases, see Figure~\ref{errLc3}. For the parameter $D$ the situation depicted in  Figure~\ref{errLD3} is opposite and the difference is increasing in $D$ which is expected as the multiplication factor is exponentially decaying in $D^{-1}.$ Finally, Figure~\ref{errLk3} suggests that the parameter $k$  seems have no substantial impact on the difference.

\end{document}